\documentclass[final]{IEEEtran}
\usepackage{cite}
\usepackage{threeparttable}
\usepackage{graphicx}
\usepackage{picinpar}
\usepackage[cmex10]{amsmath}
\usepackage{amsmath,amsfonts,amssymb}
\usepackage{subfigure}
\usepackage{algorithm}
\usepackage{algorithmic}
\usepackage{stfloats}
\usepackage{bm}
\usepackage{amsthm}
\begin{document}

\newtheorem{lemma}{Lemma}
\newtheorem{corol}{Corollary}
\newtheorem{theorem}{Theorem}
\newtheorem{proposition}{Proposition}
\newtheorem{definition}{Definition}
\newcommand{\e}{\begin{equation}}
\newcommand{\ee}{\end{equation}}
\newcommand{\eqn}{\begin{eqnarray}}
\newcommand{\eeqn}{\end{eqnarray}}

\title{\LARGE Structured Compressive Sensing Based Spatio-Temporal \\Joint Channel Estimation for FDD Massive MIMO}

\author{Zhen Gao, Linglong Dai, Wei Dai, Byonghyo Shim, and Zhaocheng Wang

\thanks{Z. Gao, L. Dai, and Z. Wang are with Department of Electronic
 Engineering, Tsinghua University, Beijing 100084, China (E-mails: gaozhen010375@foxmail.com; \{daill, zcwang\}@tsinghua.edu.cn). }

\thanks{ W. Dai is with Department of Electrical and Electronic Engineering, Imperial College London, London SW7~2AZ,~UK~(E-mail: wei.dai1@imperial.ac.uk).}
 \thanks{B. Shim is with the Department of Electrical and Computer Engineering, Seoul National University, Seoul 151-742, Korea (E-mail:bshim@snu.ac.kr).}

\thanks{This work was supported by the National Key Basic Research Program of China (Grant No. 2013CB329203),  the National Natural Science
Foundation of China (Grant Nos. 61571270 and 61201185),  the Beijing Natural Science Foundation (Grant No. 4142027), and the Foundation of Shenzhen government.} }
 %\vspace*{-3.0mm}
%}

\maketitle
%\vspace*{-16.0mm}
\begin{abstract}% ÀûÓÃÐŵÀÏà¹ØÐÔ£¿
%Massive MIMO is a promising technology for future wireless communications. However, the channel state information (CSI) acquisition in massive MIMO is a challenging task due to numerous antennas at the base station, especially for frequency division duplexing systems. Extensive experiments and studies in recent years have shown that wireless MIMO channels exhibit spatial common sparsity, which enlightens us a new perspective to solve the channel estimation problem. In this paper, we firstly provide the theoretical proof that the CSI matrix for massive MIMO over practical sparse multipath channels satisfies the asymptotical orthogonality, which indicates that sparse MIMO channels could provide favorable propagation condition. Then, we propose a bock compressive sensing (BCS) based channel estimation scheme for massive MIMO with high spectral efficiency. Specifically, at the transmitter, pilots of different antennas occupy the identical subcarriers in the frequency domain, which is different from the conventional orthogonal pilot. Accordingly, a block sparsity adaptive match pursuit algorithm is proposed to acquire CSI at the receiver. Simulation results have demonstrated that the proposed BCS based channel estimation scheme can achieve high accuracy with low pilot overhead, and it can approach the least squares method with known sparsity as the performance bound.
Massive MIMO is a promising technique for future 5G communications due to its high spectrum and energy efficiency. To realize its potential performance gain, accurate channel estimation is essential. However, due to massive number of antennas at the base station (BS), the pilot overhead required by conventional channel estimation schemes will be unaffordable, especially for frequency division duplex (FDD) massive MIMO.
To overcome this problem, we propose a structured compressive sensing (SCS)-based spatio-temporal joint channel estimation scheme to reduce the required pilot overhead, whereby the spatio-temporal common sparsity of delay-domain MIMO channels is leveraged. Particularly, we first propose the non-orthogonal pilots at the BS under the framework of CS theory to reduce the required pilot overhead. Then, an adaptive structured subspace pursuit (ASSP) algorithm at the user is proposed to jointly estimate channels associated with multiple OFDM symbols from the limited number of pilots, whereby the spatio-temporal common sparsity of MIMO channels is exploited to improve the channel estimation accuracy. Moreover, by exploiting the temporal channel correlation, we propose a space-time adaptive pilot scheme to further reduce the pilot overhead. Additionally, we discuss the proposed
channel estimation scheme in multi-cell scenario. Simulation results demonstrate that the proposed scheme can accurately estimate channels with the reduced pilot overhead, and it is capable of approaching the optimal oracle least squares estimator.
\end{abstract}
%\vspace*{-4.0mm}
\begin{IEEEkeywords}
Massive MIMO, structured compressive sensing (SCS), frequency division duplex (FDD), channel estimation.
\end{IEEEkeywords}

% \vspace*{-2.0mm}
\IEEEpeerreviewmaketitle
\section{Introduction}% ´ó¹æÄ£mimoºÜÖØÒª£¬ÐŵÀ¹À¼Æ£¬ÐŵÀ¹À¼ÆÎÊÌâ¡£
\IEEEPARstart{M}{assive} MIMO employing a large number of antennas at the base station (BS) to simultaneously serve multiple users, has recently emerged as a promising approach to realize high-throughput green wireless communications \cite{{LSMIMO2014}}. By exploiting the large number of degrees of spatial freedom, massive MIMO can boost the system capacity and energy efficiency by orders of magnitude. Therefore, massive MIMO has been widely recognized as a key enabling technique for future spectrum and energy efficient 5G communications \cite{{Overview_massive_MIMO}}.

%Although massive MIMO enjoys exciting benefits, some challenging problems have to be solved.
%One of the important issues is t
In massive MIMO systems, an accurate acquisition of the channel state information (CSI) is essential for signal detection, beamforming, resource allocation, etc. However, due to massive antennas at the BS, each user has to estimate channels associated with hundreds of transmit antennas, which results in the prohibitively high pilot overhead.
Hence, how to realize the accurate channel estimation with the affordable pilot overhead becomes a challenging problem, especially for frequency division duplex (FDD) massive MIMO systems~\cite{scaleMIMO}. To date, most of researches on massive MIMO sidestep this challenge by assuming the time division duplex (TDD) protocol, where the CSI in the uplink can be more easily acquired at the BS due to the small number of single-antenna users and the powerful processing capability of the BS, and then the channel reciprocity property can be leveraged to directly obtain the CSI in the downlink \cite{{TDD_Chen}}. However, %since we have to reuse the limited number of orthogonal pilots in adjacent cells, TDD massive MIMO suffers from the well-known problem of pilot contamination~\cite{TDD_Chen}. Moreover,
due to the calibration error of radio frequency chains and limited coherence time, the CSI acquired in the uplink may not be accurate for the downlink~\cite{{TIT_nonideal},{FDD_book}}. More importantly, compared with TDD systems, FDD systems can provide more efficient communications with low latency~\cite{TDD_FDD}, and it has dominated current cellular systems. Therefore, it is of importance to explore the challenging problem of channel estimation for FDD massive MIMO systems, which can facilitate massive MIMO to be backward compatible with current FDD dominated cellular networks.

Recently, there have been extensive studies on channel estimation for conventional small-scale FDD MIMO systems \cite{{TSP_optimal},{FP_optimal},{LTE_review},{TIT_training},{Rician},{3GPP},{channel_model_for4g}}. It has been proven that the equi-spaced and equi-power orthogonal pilots can be optimal to estimate the non-correlated Rayleigh MIMO channels for one OFDM symbol, where the required pilot overhead increases with the number of transmit antennas~\cite{{TIT_training}}. %. % , where pilots of different transmit antennas occupy the different frequency-domain subcarriers and
By exploiting the spatial correlation of MIMO channels, the pilot overhead to estimate Rician MIMO channels can be reduced \cite{Rician}. Furthermore, by exploiting the temporal channel correlation, further reduced pilot overhead can be achieved to estimate MIMO channels associated with multiple OFDM symbols \cite{{TSP_optimal},{FP_optimal}}.
Currently, orthogonal pilots have been widely used in the existing MIMO systems, where the pilot overhead is not a big issue due to the small number of transmit antennas (e.g., up to eight antennas in LTE-Advanced system)~\cite{{LTE_review},{3GPP},{channel_model_for4g}}. %, such as LTE-A, where pilots of different transmit antennas occupy different frequency-domain subcarriers, and null subcarriers are used to guarantee the orthogonality among pilots of different transmit antennashe channel estimation in MIMO systems can be easily converted to that in single-antenna systems
%Moreover, such pilot orthogonality in the frequency domain can be extended to the time domain and code domain by assuming the block fading with a certain performance loss \cite{{FP_optimal},{LTE_review}}. % MIMO due
However, this issue can be critical in massive MIMO systems due to massive number of antennas at the BS (e.g., 128 antennas or even more at the BS~\cite{Overview_massive_MIMO}).
%For instance, the pilot occupation ratio in LTE-A system with 8 transmit antennas is about 25~\% \cite{channel_model_for4g}. One can envision that for massive MIMO systems with the number of transmit antenna $M>32$, the pilot occupation ratio by using current pilot design and channel estimation scheme can be more than 100\%. Therefore, an efficient channel estimation scheme with spectral-efficiency pilot design is urgent to be resolved for FDD massive MIMO systems.
%\cite{{CS_dai},{BMSB},{CS_CE_LSMIMO},{Love}} have proposed the channel estimation schemes for downlink FDD massive MIMO.

In \cite{CS_dai}, an approach to exploit the temporal correlation
and sparsity of delay-domain
channels for the reduced pilot overhead has been proposed for FDD massive MIMO systems, but the interference cancellation of training
sequences of different transmit
antennas will be difficult when the number of transmit antennas is large. \cite{BMSB,my_cl,my_el} leveraged
the spatial correlation
and sparsity of delay-domain MIMO channels to estimate channels with the reduced pilot
overhead, but the assumption of
the known channel sparsity level at the user is unrealistic. By exploiting the spatial
channel correlation, the compressive
sensing (CS)-based channel estimation schemes were proposed in \cite{CS_CE_LSMIMO,my_TSP,Shen_CL}, but the leveraged spatial correlation can be impaired due to the non-ideal
antenna array \cite{{scaleMIMO},{TIT_nonideal}}. \cite{Love} proposed an
open-loop and closed-loop channel estimation scheme for massive MIMO, but the long-term
channel statistics perfectly known at the user can be difficult. %Additionally, \cite{} and \cite{} proposed the CS-based joint channel estimation and channel feedback for FDD massive MIMO systems.

On the other hand, for typical broadband wireless communication systems, delay-domain channels intrinsically exhibit the sparse nature due to the limited number of significant scatterers in the propagation environments and large channel delay spread~\cite{{CS_Sparse},{CS_Sparse2},{CS_dai},{GuiG},{xiaodong},{over},{my_cl2},{my_twc}}. Meanwhile, for MIMO systems with co-located antenna array at the BS, channels between one user and different transmit antennas at the BS exhibit very similar path delays due to very similar scatterers in the propagation environments, which indicates that delay-domain channels between the user and different transmit antennas at the BS share the common sparsity when the aperture of the antenna array is not very large \cite{{scaleMIMO},{outdoor}}. Moreover, since the path delays vary much slower than the path gains due to the temporal channel correlation, such sparsity is almost unchanged during the coherence time \cite{{Channel_corr}}. In this paper, such channel properties of MIMO channels are referred to as the \emph{spatio-temporal common sparsity}, which is usually not considered in most of current work.

In this paper, by exploiting the spatio-temporal common sparsity of delay-domain MIMO channels, we propose a structured compressive sensing (SCS)-based spatio-temporal joint channel estimation scheme with significantly reduced pilot overhead for FDD massive MIMO systems. %Particularly, %the contributions of this paper are listed as below:
%For each user, by exploiting the i, wre listed as
%\begin{itemize}
  %\item
Specifically, at the BS, we propose a non-orthogonal pilot scheme under the framework of CS theory, which is essentially different from the widely used orthogonal pilots under the framework of classical Nyquist sampling theorem. Compared with conventional orthogonal pilots, the proposed non-orthogonal pilot scheme can substantially reduce the required pilot overhead for channel estimation.
At the user side, we propose an adaptive structured subspace pursuit (ASSP) algorithm for channel estimation, whereby the spatio-temporal common sparsity of delay-domain MIMO channels is leveraged to improve the channel estimation performance from the limited number of pilots. %{\color{red}Compared with the model-based SP algorithm only exploiting the structured sparsity  \cite{model}, the proposed ASSP algorithm}
%{\color{red}At the user side, we propose an adaptive structured subspace pursuit (ASSP) algorithm developed from the classical subspace pursuit (SP) algorithm \cite{SP} for channel estimation}, whereby the spatio-temporal common sparsity of delay-domain MIMO channels is leveraged to improve the channel estimation performance from the limited number of pilots. {\color{red}Compared with the model-based SP algorithm only exploiting the structured sparsity  \cite{model}, the proposed ASSP algorithm}
Furthermore, by leveraging the temporal channel correlation, we propose a space-time adaptive pilot scheme to realize the accurate channel estimation with further reduced pilot overhead, where the specific pilot signals should consider the geometry of antenna array at the BS and the mobility of served users. %Meanwhile, it can reduce the computational complexity of channel estimation scheme at the user.
Additionally, we further extend the proposed channel estimation scheme from the single-cell scenario to the multi-cell
scenario. Finally, simulation results verify that the proposed scheme outperforms its conventional counterparts with reduced pilot overhead, where the performance of the SCS-based channel estimation scheme approaches that of the oracle least squares (LS) estimator.%, and the parametric channel feedback scheme only suffers from a negligible performance loss compared to the complete channel feedback scheme.%the sum-rate of with the low pilot overhead.
% \end{itemize}
%Simulation results verify that the realistic sparse MIMO channels could provide the favorable propagation conditions in typical massive MIMO scenarios, and the proposed ASSP algorithm can reliably acquire CSI in the downlink with low pilot overhead.

%In MIMO systems with small to moderate scale, the identification of MIMO channels can be obtained by the conventional orthogonal pilot based channel estimation schemes, where valid pilots associated with different antennas occupy different frequency-domain subcarriers, and zero pilots are exploited to avoid the mutual interferences of pilots from different antennas \cite{FP}. In this way, the channel estimation in MIMO systems can be easily converted to that in single antenna systems. However, such schemes suffers from high pilot overhead when the number of antennas become large since the pilot overhead in conventional schemes heavily depends on the maximum delay spread of channels and increases with the number of antennas. To solve this problem, several previous work endeavors to reduce the pilot overhead \cite{{FP_optimal},{LTE_review},{broad}}, however, these schemes are confined to the trade-off between the accuracy of channel estimation and pilot overhead. Moreover, since the number of antennas at BS in massive MIMO can be up to dozens even hundreds, efforts to reduce the pilot overhead in conventional MIMO channel estimation schemes can be a drop in the bucket.

% ÔÚÕâôһ¸ö±³¾°Ï£¬

The rest of the paper is organized as follows. Section II illustrates the spatio-temporal common sparsity of delay-domain MIMO channels. In Section III, the proposed SCS-based spatio-temporal joint channel estimation scheme is discussed in detail. In Section IV, we provide the performance analysis. Section V shows the simulation results. Finally, Section VI concludes this paper.

{\it Notation: }Boldface lower and upper-case symbols represent column vectors and matrices, respectively. The operator $\circ $ represents the Hadamard product, $\lfloor \cdot \rfloor$ denotes the integer floor operator, and $\rm{diag}\{\bf{x}\}$ is a diagonal matrix with elements of $\mathbf{x}$ on its diagonal. The matrix inversion, transpose, and Hermitian transpose operations are denoted by $(\cdot )^{-1}$, $(\cdot )^{ \rm{T}}$, and $(\cdot )^{\rm{H}}$, respectively, while $(\cdot )^{\dag}$ denotes the Moore-Penrose matrix inversion. $| \cdot |_c$ denotes the cardinality of a set, the $l_2$-norm operation and Frobenius-norm operation are given by $\|\cdot\|_2$ and $\|\cdot\|_F$, respectively. ${\Omega ^c}$ denotes the complementary set of the set $\Omega $. $\rm{Tr}\{\cdot\}$ is the trace of a matrix. $\left\langle \cdot, \cdot \right\rangle $ is the Frobenius inner product, and $\left\langle {{\bf{A}}{\rm{,}}{\bf{B}}} \right\rangle {\rm{ = Tr\{ }}{{\bf{A}}^{\rm{H}}}{\bf{B}}{\rm{\} }}$. Finally, $\mathbf{\Phi}^{(l)}$ denotes the $l$th column vector of the matrix $\mathbf{\Phi}$. %Finally, $ \mathbf{x}_{\Gamma}$ denotes the entries of $\mathbf{x}$ defined in the set $\Gamma$, while $\mathbf{H}_{\Gamma}$ denotes a sub-matrix of $\mathbf{H}$ with indices of columns defined by the set~$\Gamma$.% and $\rho_n \left(\cdot  \right)$ denotes the spectral norm of a matrix.

%\vspace*{-2.0mm}
\section{Spatio-Temporal Common Sparsity of Delay-Domain} \label{SCS}%the majority of channel gains end up being either zero or very small in typical wireless %communication systems
Extensive experimental studies have shown that wireless broadband channels exhibit
the \emph{sparsity} in the
delay domain. This is caused by the fact that the number of multipath dominating
the majority of channel energy
is small due to the limited number of significant scatterers in the wireless signal
propagation environments,
while the channel delay spread can be large due to the large difference between the
time of arrival (ToA) of
the earliest multipath and the ToA of the latest multipath \cite{{CS_Sparse},{CS_Sparse2},{CS_dai},{GuiG},{xiaodong},{over},{my_cl2},{my_twc}}.
Specifically, in the downlink, the delay-domain channel impulse response (CIR)
between the $m$th transmit
antenna at the BS and one user can be expressed~as
\begin{equation}\label{equ:CIR}
{{\bf{h}}_{m,r}} = {\left[ {{h_{m,r}}[1],{h_{m,r}}[2], \cdots ,{h_{m,r}}[L ]} \right]^{\rm{T}}},1 \le m \le M,
\end{equation}
where $r$ is the index of the OFDM symbol in the time domain, $L$ is the equivalent channel length, $D_{m,r}  = \text{supp}\{{{\bf{h}}_{m,r}}\}=\left\{ {l :\left| {{h_{m,r}}[l]} \right| > p_{\rm{th}}}, 1\le l\le L \right\}$ is the support set of ${{\bf{h}}_{m,r}}$, and $p_{\rm{th}}$ is the
noise floor according to~\cite{th}. The sparsity level of wireless
channels is denoted as ${P_{m,r}} = \left| {{D_{m,r}}} \right|_c$, and
we have $P_{m,r} \ll L$ due to the sparse nature of delay-domain
channels~\cite{{CS_Sparse},{CS_Sparse2},{CS_dai},{GuiG}}\footnote{The sparse delay-domain channels may exhibit the power leakage due to the non-integer normalized path delays. To solve this issue, there have been off-the-shelf techniques to mitigate the power leakage \cite{{xiaodong},{over}}. For convenience, we consider the sparse channel model in the equivalent discrete-time baseband widely used in CS-based channel estimation~\cite{{CS_Sparse},{CS_Sparse2},{CS_dai},{GuiG}}.}. %Therefore, $S_{m}$ is also referred as the (effectively) sparsity of wireless channels.

%Furthermore, practical wireless channels also exhibit temporal correlations even when they are fast time-varying \cite{IT}. It has been found that the path delays usually vary much slower than the path gains \cite{TDS_TSP}. In other words, although the path gains can be varying significantly from one OFDM symbol to another, the path delays almost remain unchanged during several successive OFDM symbols. This is because that the coherence time of path gains over time-varying channels, denoted by $T_{\text{gain}}$, is inversely proportional to the system carrier frequency, while the duration for path delay variation, denoted by $T_{\text{delay}}$, is inversely proportional to the system bandwidth. Taking the LTE-A working at carrier of gigahertz frequency band with a signal bandwidth of megahertz for instance, the path delays vary at a rate that is about 100 times slower than that of the path gains. That is to say, in the period of $T_{\text{delay}}$, there exists
%\begin{equation}\label{equ:temporal}
%D_{m,r}  = D_{m,r + 1}  =  \cdots  = D_{m,r + R - 1},1 \le m \le M,
%\end{equation}
%where the path delays are nearly unchanged during $R$ OFDM symbols. Therefore, such temporal correlation of wireless channels is also referred as the temporal common sparsity of wireless channels.

\begin{figure}[!tbp]
     \centering
    % \vspace*{-10mm}
     \includegraphics[width=6.5cm, keepaspectratio]
     {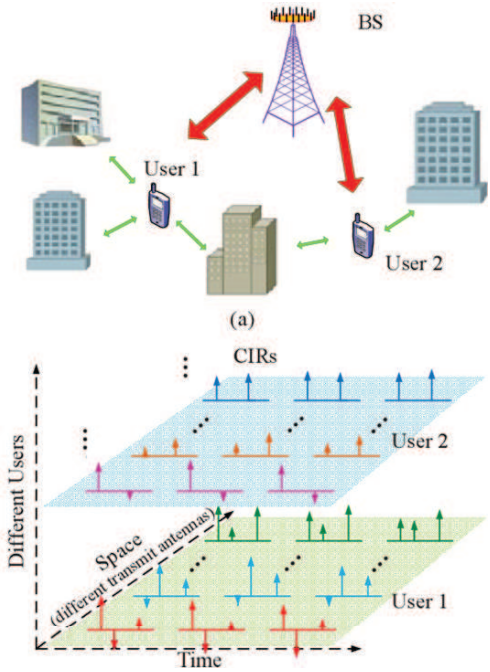}
      \vspace*{-2mm}
    \caption{Spatio-temporal common sparsity of delay-domain MIMO channels: (a) Wireless channels exhibit the sparse nature due to the limited number of scatterers; (b) Delay-domain MIMO channels between the co-located antenna array and one user exhibit the spatio-temporal common sparsity.}%; (b): Proposed superimposed pilot schemewhere horizontal axis is path delays, and vertical %axis is delay gains
     \label{fig:CIR}
      \vspace*{-2.0mm}
\end{figure}

Moreover, there are measurements showing that CIRs between different transmit antennas and one user exhibit very similar path delays \cite{{scaleMIMO},{outdoor}}. The reason is that, in typical massive MIMO geometry, the scale of the compact antenna array at the BS is relatively small compared with the large signal transmission distance, and channels associated with different transmit-receive antenna pairs share the common scatterers. Therefore, the sparsity patterns of CIRs of different transmit-receive antenna pairs have a large overlap. Moreover, for MIMO systems with not very large $M$, these CIRs can share the common sparse pattern~\cite{{BMSB},{scaleMIMO},{outdoor}}, i.e.,
\begin{equation}\label{equ:temporal}
D_{1,r}  = D_{2,r}  =  \cdots  = D_{M,r},
\end{equation}
which is referred to as the \emph{spatial common sparsity} of wireless MIMO channels. For example, we consider the LTE-Advanced system working at a carrier frequency of $f_c=2~\rm{GHz}$ with a signal bandwidth of $f_s=10~\rm{MHz}$, and the uniform linear array (ULA) with the antenna spacing of half-wavelength. For two transmit antennas with the distance of 8 half-wavelengths, their maximum difference of path delays from the common scatterer is $\frac{{{8 \lambda}/2}}{{{c}}} = 4/f_c=0.002~\mu s$, which is negligible compared with the system sample period $T_s=1/f_s=0.1~\mu s$, where $\lambda$ and $c$ are the wavelength and the velocity of light, respectively.
It should be pointed out that the path gains of different transmit-receive antenna pairs from the same scatterer can be different or even uncorrelated due to the non-isotropic antennas\footnote{For practical massive MIMO systems, different antennas at the BS with different directivities can destroy the spatial correlation of path gains over different transmit-receive pairs from the same scatterer and improve the system capacity \cite{scaleMIMO}. However, this spatial channel correlation is usually exploited in conventional channel estimation schemes for reduced pilot overhead, which can be unrealistic.} \cite{TIT_nonideal}.
%Furthermore, for typical wireless communication systems, the path delay difference from the same scatterers is far less than the system sampling period~\cite{BMSB}.

Finally, practical wireless channels also exhibit the temporal correlation even in fast time-varying scenarios \cite{{Channel_corr}}. It has been demonstrated that the path delays usually vary much slower than the path gains \cite{{Channel_corr}}. In other words, although the path gains can vary significantly from one OFDM symbol to another, the path delays remain almost unchanged during several successive OFDM symbols. This is due to the fact that the coherence time of path gains over time-varying channels is inversely proportional to the system carrier frequency, while the duration for path delay variation is inversely proportional to the system bandwidth \cite{{Channel_corr}}. For example, in the LTE-Advanced system with $f_c=2~\rm{GHz}$ and $f_s=10~\rm{MHz}$, the path delays vary at a rate that is about several hundred times slower than that of the path gains \cite{CS_dai}. That is to say, during the coherence time of path delays, CIRs associated with $R$ successive OFDM symbols have the common sparsity due to the almost unchanged path delays, i.e.,
\begin{equation}\label{equ:temporal}
D_{m,r}  = D_{m,r + 1}  =  \cdots  = D_{m,r + R - 1},1 \le m \le M.
\end{equation}
This temporal correlation of wireless channels is also referred to as the \emph{temporal common sparsity} of wireless channels in this paper.

The spatial and temporal channel correlations discussed above are jointly referred to as the \emph{spatio-temporal common sparsity} of delay-domain MIMO channels, which can be illustrated in~Fig.~\ref{fig:CIR}. This channel property is usually not considered in existing channel estimation schemes. In this paper, we will exploit this channel property to overcome the challenging problem of channel estimation for FDD massive MIMO.% as will be shown in Section III.

\begin{figure}[!tbp]
     \centering
     %\vspace*{-10.0mm}
     \includegraphics[width=9.2cm, keepaspectratio]
     {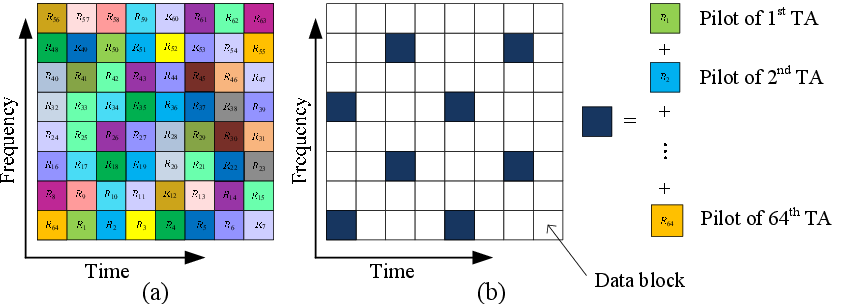}
           \vspace*{-3.0mm}
    \caption{Pilot designs for massive MIMO with $M=64$ in one time-frequency resource block. (a) Conventional orthogonal pilot design; (b) Proposed non-orthogonal pilot design.}%; (b): Proposed superimposed pilot schemewhere horizontal axis is path delays, and vertical %axis is delay gains
     \label{fig:pilot2}
      \vspace*{-2.0mm}
\end{figure}

%approaches to develop%inspires us
%to exploit inherent structures of delay-domain MIMO channels to achieve a reliable and accurate acquisition of CSI with low pilot overhead.
%\vspace*{-3.0mm}
\section{Proposed SCS-Based Spatio-Temporal Joint Channel Estimation Scheme}
%In this section, the SCS based spatio-temporal joint channel estimation scheme is proposed for FDD massive MIMO. First, we propose the frequency-domain non-orthogonal pilots at the BS to reduce the pilot overhead for channel estimation. Accordingly, we propose the ASSP algorithm at the user to estimate channel from the limited received pilot, whereby the spatio-temporal common sparsity of MIMO channels is leveraged for enhanced channel estimation performance. Moreover, we propose the parametric channel feedback scheme, which can achieve the accurate CSI at the BS with much reduced channel feedback overhead thanks to the sparsity of channels. Finally, we provide the space-time adaptive pilot scheme, which can further improve the spectrum-energy efficiency for users with different mobile~speeds.

In this section, the SCS-based spatio-temporal joint channel estimation scheme is proposed for FDD massive MIMO. First, we propose the non-orthogonal pilot scheme at the BS to reduce the pilot overhead. Then, we propose the ASSP algorithm at the user for reliable channel estimation. Moreover, we propose the space-time adaptive pilot scheme for further reduction of the pilot overhead. Finally, we briefly discuss the proposed channel estimation scheme extended to multi-cell scenario.

\vspace*{-3.0mm}
\subsection{Non-Orthogonal Pilot Scheme at the BS} \label{pilot}

The design of conventional orthogonal pilots is based on the framework of classical Nyquist sampling theorem, and this design has been widely used in the existing MIMO systems. The orthogonal pilots can be illustrated in Fig.~\ref{fig:pilot2} (a), where pilots associated with different transmit antennas occupy the different subcarriers. For massive MIMO systems with hundreds of transmit antennas, such orthogonal pilots will suffer from the prohibitively high pilot overhead.%, which can consume much spectrum and energy resources.

In contrast, the design of the proposed non-orthogonal pilot scheme, as shown in Fig. \ref{fig:pilot2} (b), is based on CS theory, and it allows pilots of different transmit antennas to occupy the completely same subcarriers. By leveraging the sparse nature of channels, the pilots used for channel estimation can be reduced substantially.%, which indicates that the spectrum and energy resources for effective data transmission can be substantially increased.

For the proposed non-orthogonal pilot scheme, we first consider the MIMO channel estimation for one OFDM symbol as an example. Particularly, we denote the index set of subcarriers allocated to pilots as $\xi$, which is uniquely selected from the set of $\left\{ {1,2, \cdots ,N} \right\}$ and identical for all transmit antennas. Here ${N_p} = \left| \xi  \right|_c$ is the number of pilot subcarriers in one OFDM symbol, and {$N$ is the number of subcarriers in one OFDM symbol}. Moreover, we denote the pilot sequence of the $m$th transmit antenna as ${{\bf{p}}_m}\in \mathbb{C}^{N_p\times 1}$. The specific pilot design $\xi$ and $\left\{ {{{\bf{p}}_m}} \right\}_{m = 1}^M$ will be detailed in Section \ref{pilot_design} .
\vspace*{-2.0mm}
\subsection{SCS-Based Channel Estimation at the User} \label{CE}
At the user, after the removal of the guard interval and discrete Fourier transformation (DFT), the received pilot sequence ${{\bf{y}}_r}\in \mathbb{C}^{N_p \times 1}$ of the $r$th OFDM symbol can be expressed as
\begin{equation}\label{equ:re_pilot}
\begin{array}{l}
{{\bf{y}}}_r = \sum\limits_{m = 1}^M {{\rm{diag}}\{ {{\bf{p}}_{m}}\} {{\left. {\bf{F}} \right|}_\xi }\left[ {\begin{array}{*{20}{l}}
{{{\bf{h}}_{m,r}}}\\
{{{\bf{0}}_{(N - L) \times 1}}}
\end{array}} \right]}  + {\bf{w}}_r\\
~~~ = \sum\limits_{m = 1}^M {{{\bf{P}}_m}{{\left. {{{\bf{F}}_L}} \right|}_\xi }{{\bf{h}}_{m,r}}}  + {\bf{w}}_r = \sum\limits_{m = 1}^M {{\bf{\Phi}}_m {{\bf{h}}_{m,r}}}  + {\bf{w}}_r,
\end{array}
\end{equation}
where ${{\bf{P}}_m}{\rm{ = diag}}\{ {{\bf{p}}_m}\} $, ${\bf{F}}\in \mathbb{C}^{N\times N}$ is a DFT matrix, ${{\bf{F}}_L}\in \mathbb{C}^{N\times L}$ is a partial DFT matrix consisted of the first $L$ columns of ${\bf{F}}$, ${\left. {{{\bf{F}}}} \right|_\xi }\in \mathbb{C}^{N_p\times N}$ and ${\left. {{{\bf{F}}_L}} \right|_\xi }\in \mathbb{C}^{N_p\times L}$ are the sub-matrices by selecting the rows of ${{{\bf{F}}}}$ and ${{{\bf{F}}_L}}$ according to $\xi$, respectively, ${\bf{w}}_r\in \mathbb{C}^{N_p\times 1}$ is the additive white Gaussian noise (AWGN) vector in the $r$th OFDM symbol, and ${{\bf{\Phi }}_m} = {{\bf{P}}_m}{\left. {{{\bf{F}}_L}} \right|_\xi }$. Moreover, (\ref{equ:re_pilot}) can be rewritten in a more compact form as
\vspace*{-2.0mm}
\begin{equation}\label{equ:compact}
{\bf{y}}_r = {{\bf{\Phi}} {\bf{\tilde h}}_r} + {\bf{w}}_r,
\end{equation}
where ${\bf{\Phi }} = \left[ {{{\bf{\Phi }}_1},{{\bf{\Phi }}_2}, \cdots ,{{\bf{\Phi }}_M}} \right]\in \mathbb{C}^{{N_p} \times ML}$, and ${\bf{\tilde h}}_r = {[{\bf{h}}_{1,r}^{\rm{T}},{\bf{h}}_{2,r}^{\rm{T}}, \cdots ,{\bf{h}}_{M,r}^{\rm{T}}]^{\rm{T}}}\in \mathbb{C}^{ ML \times 1}$ is an aggregate CIR vector.

For massive MIMO systems, we usually have ${N_p} \ll ML$ due to the large number of transmit antennas $M$ and the limited number of pilots $N_p$. This indicates that we cannot reliably estimate ${\bf{\tilde h}}_r$ from ${\bf{y}}_r$ using conventional channel estimation schemes, since (\ref{equ:compact}) is an under-determined system. However, the observation that ${\bf{\tilde h}}_r$ is a sparse signal due to the sparsity of $\{ {{\bf{h}}_{m,r}}\} _{m = 1}^M$ inspires us to estimate the sparse signal ${\bf{\tilde h}}_r$ of high dimension from the received pilot sequence ${\bf{y}}_r$ of low dimension under the framework of CS theory \cite{STR_CS}. Moreover, the inherently spatial common sparsity of wireless MIMO channels can be also exploited for performance enhancement. Specifically, we rearrange the aggregate CIR vector ${\bf{\tilde h}}_r$ to obtain the equivalent CIR vector ${\bf{\tilde d}}_r$ as
\begin{equation}\label{equ:d}
{{\bf{\tilde d}}}_r = {[{\bf{d}}_{1,r}^{\rm{T}},{\bf{d}}_{2,r}^{\rm{T}}, \cdots ,{\bf{d}}_{L,r}^{\rm{T}}]^{\rm{T}}}\in \mathbb{C}^{{ML} \times 1},
\end{equation}
where ${\bf{d}}_{l,r}^{\rm{}} = \left[ {{h_{1,r}}[l],{h_{2,r}}[l], \cdots ,{h_{M,r}}[l]} \right]^{\rm{T}}$ for $1\le l\le L$. Similarly, ${\bf{\Phi }}$ can be rearranged as ${\bf{\Psi }}$, i.e.,
\begin{equation}\label{equ:kesai}
{\bf{\Psi }} = \left[ {{{\bf{\Psi }}_1},{{\bf{\Psi }}_2},\cdots ,{{\bf{\Psi }}_{L}}} \right]\in \mathbb{C}^{{N_p} \times ML},
\end{equation}
where ${{\bf{\Psi }}_l} = \left[ {{{ {{{\bf{\Phi }}_1^{(l)}}}}},{{ {{{\bf{\Phi }}_2^{(l)}}} }}, \cdots ,{{ {{{\bf{\Phi }}_M^{(l)}}} }}} \right] = \left[ {{{\bm{\psi }}_{1,l}},{{\bm{\psi }}_{2,l}}, \cdots ,{{\bm{\psi }}_{M,l}}} \right]\in \mathbb{C}^{{N_p} \times M}$. In this way, (\ref{equ:compact}) can be reformulated as
\vspace*{-2.0mm}
\begin{equation}\label{equ:compact2}
{{\bf{y}}}_r = {\bf{\Psi }}{{\bf{\tilde d}}}_r + {{\bf{w}}}_r.
\vspace*{-2.0mm}
\end{equation}
From (\ref{equ:compact2}), it can be observed that due to the spatial common sparsity of wireless MIMO channels, the equivalent CIR vector ${\bf{\tilde d}}_r$ exhibits the structured sparsity~\cite{STR_CS}.%, which motivates us to exploit the theory of block-sparse CS developed from the classical CS theory to simultaneously acquire CSI associated with different transmit antennas~\cite{STR_CS}.

Furthermore, the temporal correlation of wireless channels indicates that such spatial common sparsity in MIMO systems remains virtually unchanged over $R$ successive OFDM symbols, where $R$ is determined by the coherence time of the path delays~\cite{CS_dai}. Hence, wireless MIMO channels exhibit the spatio-temporal common sparsity during $R$ successive OFDM symbols. Considering~(\ref{equ:compact2}) during $R$ adjacent OFDM symbols with the same pilot pattern, we have%, which is quite common in practice \cite{TSE},
\vspace*{-2.0mm}
\begin{equation}\label{equ:common}
{\bf{Y}} = {\bf{\Psi D}} + {\bf{W}},
\vspace*{-2.0mm}
\end{equation}
where ${\bf{Y}}\!=\!\left[ {{{\bf{y}}_r},\!{{\bf{y}}_{r + 1}}, \!\cdots\! ,\!{{\bf{y}}_{r + R - 1}}} \right]\!\in \!\mathbb{C}^{N_p\times R}$ is the measurement matrix, ${\bf{D}}\! =\! \left[ {{{{\bf{\tilde d}}}_r},\!{{{\bf{\tilde d}}}_{r + 1}},\! \cdots \! , \! {{{\bf{\tilde d}}}_{r + R - 1}}} \right]\!\in \! \mathbb{C}^{ ML \times R} $~is the equivalent CIR matrix, and ${\bf{W}} = \left[ {{{\bf{w}}_r},{{\bf{w}}_{r + 1}}, \cdots ,{{\bf{w}}_{r + R - 1}}} \right]\in \mathbb{C}^{ N_p\times R} $~is the AWGN matrix.
It should be pointed out that ${\bf{D}}$ can be expressed as
\vspace*{-2.0mm}
\begin{equation}\label{equ:common3}
{\bf{D}} = {\rm{ }}{[{\bf{D}}_1^{\rm{T}},{\bf{D}}_2^{\rm{T}}, \cdots ,{\bf{D}}_L^{\rm{T}}]^{\rm{T}}},
\vspace*{-2.0mm}
\end{equation}
where ${{{\bf{D}}_l}}$ for $1\le l\le L$ has the size of $M\times R$, and the $m$th row and $r $th column element of ${{{\bf{D}}_l}}$ is the channel gain of the $l$th path delay associated with the $m$th transmit antenna in the $r$th OFDM symbol.

It is clear that the equivalent CIR matrix $\bf{D}$ in (\ref{equ:common3}) exhibits the structured sparsity due to the spatio-temporal common sparsity of wireless MIMO channels, and this intrinsic sparsity in $\bf{D}$ can be exploited for better channel estimation performance. In this way, we can jointly estimate channels associated with $M$ transmit antennas in $R$ OFDM symbols by jointly processing the received pilots of $R$ OFDM symbols.

%The spatial common sparsity of wireless MIMO channels as addressed above motivates us to utilize the multiple measurement vectors model (MMV) based CS \cite{STR_CS}, which is developed from the conventional single measurement vector (SMV) model based CS. In this way, we can jointly estimate MIMO channels of $R$ adjacent OFDM symbols.

By exploiting the structured sparsity of ${{\bf{D}}}$ in (\ref{equ:common}), we propose the ASSP algorithm as described in {Algorithm 1} to estimate channels for massive MIMO systems. Developed from the classical subspace pursuit (SP) algorithm \cite{SP}, the proposed ASSP algorithm exploits the structured sparsity of $\bf{D}$ for further improved sparse signal recovery performance.% Moreover, ASSP algorithm can adaptively acquire the channel sparsity level, which is usually required by conventional CS algorithms as the priori information, including the classical SP algorithm.

\begin{algorithm}[tp]
\begin{small}
%\algsetup{linenosize=\scriptsize}\scriptsize
\renewcommand{\algorithmicrequire}{\textbf{Input:}}
\renewcommand\algorithmicensure {\textbf{Output:} }
\caption{Proposed ASSP Algorithm.}
\label{alg:Framwork} % Alg 1
\begin{algorithmic}[1]
\REQUIRE
Noisy measurement matrix ${\bf{Y}}$ and sensing matrix ${\bf{\Psi }}$.% in (\ref{equ:compact2}).%, common sparsity level $K$, maximum channel length $L$, number of transmit antenna $M$, adjacent OFDM symbol $R$.
\ENSURE
The estimation of channels $\left\{ {{{\bf{ h}}_{m,t}}} \right\}_{m = 1,t = r}^{m = M,t = r + R - 1}$. \\
%\Inization ~~\\                          %Ëã·¨µÄÊäÈë²ÎÊý£ºInitialization

${\kern -7pt}$$ \bullet $ \textbf{Step 1} (\emph{Initialization}) The initial channel sparsity level $s= 1$, the iterative index $k= 1$, the support set $\Omega^{k-1}  = \emptyset$, and the residual matrices ${\bf{R}}^{k-1}  =  \bf{Y}$ and ${\left\|{\bf{R}}_{s-1} \right\|_F} =  + \inf $.\\
%\STATE $ {\bf{C}} =\bf{0}$;  $\{$Initialization$\}$
%\STATE $\Omega^0  = \emptyset$;  $\{$Empty$\}$
%\STATE ${\bf{R}}^0  = \bf{Y}$;  $\{$Initial residual$\}$
%\STATE ${{\bf{R}}_{{\rm{last}}}} =  + \inf $;  $\{$Initial residual of last stage$\}$
%\STATE $s = 1$;  $\{$Step size$\}$
%\STATE ${\cal{T}} = s$;  $\{$Size of the finalist in the first stage$\}$
%\STATE $i = 1$;  $\{$Iteration index$\}$
%\STATE $j =1$;  $\{$Stage index$\}$
${\kern -7pt}$$ \bullet $ \textbf{Step 2} (\emph{Solve the Structured Sparse Matrix $\bf{D}$ to (9)}) \\
  $ \textbf{repreat}$   \\
% ${\bf{A}} = {{\bf{\Psi }}^{\rm{H}}}{{\bf{R}}^{k - 1}}$;  $\{$Correlation$\}$
$ \text{1.}$ (\emph{Correlation}) ~~~~~~${\kern 2pt}$${\bf{Z}} = {{\bf{\Psi }}^{\rm{H}}}{{\bf{R}}^{k - 1}}$;\\
$ \text{2.}$ (\emph{Support Estimate}) $\tilde \Omega^{'k}  = {\Omega ^{k - 1}} \cup {\Pi^s}\left( {\left\{ {{{\left\| {{{\bf{Z}}_l}} \right\|}_F}} \right\}_{l = 1}^{L }} \right)$;\\%$\Gamma  = \mathop {{\rm{max}}}\limits_l \{ {\left\| {{{\bf{A}}_l}} \right\|_F},\cal{T}\}$;  ${\Pi }={\Omega ^{i - 1}} \cup {\Gamma}$;\\
%$ \text{2.}$ (\emph{Support Merge})~~ ${\Pi }={\Omega ^{i - 1}} \cup {\Gamma}$; \\
$ \text{3.}$ (\emph{Support Pruning})${\kern 1pt}$ ${\bf{{{\bf{\mathord{\buildrel{\lower3pt\hbox{$\scriptscriptstyle\smile$}}
\over D}}}}}}_{\tilde \Omega^{'k}} = {\bf{\Psi }}_{{\tilde \Omega }^{'k}}^\dag {\bf{Y}}$;  ${\bf{{{\bf{\mathord{\buildrel{\lower3pt\hbox{$\scriptscriptstyle\smile$}}
\over D}}}}}}_{(\tilde \Omega^{'k})^c} = {\bf{0}}$;\\
~~~~~~~~~~~~~~~~~~~~~~~~~${\kern +0.3pt}$~${\tilde \Omega^{k}} = {\Pi^s}\left( {\left\{ {{{\left\| {{{{\bf{\mathord{\buildrel{\lower3pt\hbox{$\scriptscriptstyle\smile$}}
\over D}}}}_l}} \right\|}_F}} \right\}_{l =1}^{L }} \right)$;\\
$ \text{4.}$ (\emph{Matrix Estimate}) ${\kern 1pt}$ ${\bf{{{{\mathord{\buildrel{\lower3pt\hbox{$\scriptscriptstyle\smile$}}
\over D}}}}}}_{\tilde \Omega^{k}} = {\bf{\Psi }}_{{\tilde \Omega^{k} }}^\dag {\bf{Y}}$; ${\bf{{{{\mathord{\buildrel{\lower3pt\hbox{$\scriptscriptstyle\smile$}}
\over D}}}}}}_{(\tilde \Omega^{k})^c} = {\bf{0 }}$;\\
$ \text{5.}$ (\emph{Residue Update})${\kern 5pt}$  ${\bf{R}}^{k} = {\bf{Y}} - {{\bf{\Psi }} }{\bf{\mathord{\buildrel{\lower3pt\hbox{$\scriptscriptstyle\smile$}}
\over D}}}$;\\
$ \text{6.}$ (\emph{Matrix Update})${\kern 5pt}$~~${{\bf{\mathord{\buildrel{\lower3pt\hbox{$\scriptscriptstyle\smile$}}
\over D}}}^{k}} ={\bf{\mathord{\buildrel{\lower3pt\hbox{$\scriptscriptstyle\smile$}}
\over D}}}$; \\
%\STATE ${\Pi }={\Omega ^{i - 1}} \cup {\Gamma}$;  $\{$Make candidate list$\}$
%\STATE ${\bf{B}}_{\Pi} = {\bf{\Psi }}_{{\Pi }}^\dag {\bf{Y}}$;  $\{$Least squares$\}$
%\STATE ${\Omega} = \mathop {{\rm{max}}}\limits_l \{ {\left\| {{{\bf{B}}_l}} \right\|_F},{\cal{T}}\}$;  $\{$Final test$\}$
%\STATE ${\bf{C}}_{\Omega} = {\bf{\Psi }}_{{\Omega }}^\dag {\bf{Y}}$;  $\{$Least squares$\}$
%\STATE ${\bf{R}} = {\bf{Y}} - {{\bf{\Psi }} }{\bf{C}}$;  $\{$Compute residual$\}$
%\STATE ${l_{{\rm{min}}}} = \mathop {{\rm{min}}}\limits_l \{ {\left\| {{\bf{C}}_l} \right\|_F},l \in \Omega \} $;%  $\{$Compute residual$\}$
 ~~$ \textbf{if}$  {${\left\| {{{\bf{R}}^{k - 1}}} \right\|_F} > {\left\| {\bf{R}}^{k} \right\|_F}$}\\
$ \text{7.}$ (\emph{Iteration with Fixed Sparsity Level}) ${\Omega}^k = {\tilde \Omega^{k}}$; $k  = k+1$; \\
%    \STATE ${\Omega}^i = {\Omega}$;  $\{$Update the finalist$\}$
%    \STATE ${{\bf{R}}^i} = {\bf{R}}$;  $\{$Update residual$\}$
%    \STATE $i  = i+1$;  $\{$Update the iteration index$\}$
 ~~$ \textbf{else}$\\
$ \text{8.}$  (\emph{Update Sparsity Level}) ${{\bf{\mathord{\buildrel{\lower3pt\hbox{$\scriptscriptstyle\smile$}}
\over D}}}_{s}} ={{\bf{\mathord{\buildrel{\lower3pt\hbox{$\scriptscriptstyle\smile$}}
\over D}}}}^{k-1}$; ${{\bf{R}}_{{s}}} ={\bf{R}}^{k-1}$; \\
~~~~~~~~~~~~~~~~~~~~~~~~~~~~~~~~$\Omega_s ={\Omega}^{k-1} $; $s= s+1$;\\ %$k  = k+1$; \\
% ${\Omega}_s = {\tilde  \Omega^{k}}$;
%    \STATE $j = j+1$;  $\{$Update the stage index$\}$
%    \STATE ${\cal{T}}  = j\times s$;  $\{$Update the size of finalist$\}$
%    \STATE ${\bf{C}_{\rm{last}}} ={\bf{C}}$;  $\{$Store sparse matrix of the last stage$\}$
%    \STATE ${\bf{R}_{\rm{last}}} ={\bf{R}}$;  $\{$Store residual of the last stage$\}$
 ~~$ \textbf{end if}$\\

     $ \textbf{until}$ {stopping criteria are met}  %\textbf{do}$

%\REPEAT%{${\left\| {{{{\bf{V}}}_k}} \right\|_F}  <  {\left\|( {{{\bf{V}}_{k - 1}}}) \right\|_F}$,}%$k \le K$
%\WHILE

${\kern -7pt}$$ \bullet $ \textbf{Step 3} (\emph{Obtain Channels})  $\widehat {\bf{D}}  = {{\bf{\mathord{\buildrel{\lower3pt\hbox{$\scriptscriptstyle\smile$}}
\over D}}}_{s-1}}$ and obtain the estimation of channels $\left\{ {{{\bf{ h}}_{m,t}}} \right\}_{m = 1,t = r}^{m = M,t = r + R - 1}$ according to (4)-(9).

\end{algorithmic}
\end{small}
\end{algorithm}

For {Algorithm 1}, some notations should be further detailed. First, both ${\bf{Z}}\in \mathbb{C}^{{ML} \times R}$ and ${\bf{\mathord{\buildrel{\lower3pt\hbox{$\scriptscriptstyle\smile$}}
\over D}}}\in \mathbb{C}^{{ML} \times R}$ are consisted of $L$ sub-matrices with the equal size of ${{M} \times R}$, i.e., ${\bf{Z}} = {[{\bf{Z}}{_1^{\rm{T}}},{\bf{Z}}{_2^{\rm{T}}}, \cdots ,{\bf{Z}}{_{L}^{\rm{T}}}]^{\rm{T}}}$ and ${\bf{\mathord{\buildrel{\lower3pt\hbox{$\scriptscriptstyle\smile$}}
\over D}}} = {[{\bf{\mathord{\buildrel{\lower3pt\hbox{$\scriptscriptstyle\smile$}}
\over D}}}{_1^{\rm{T}}},{\bf{\mathord{\buildrel{\lower3pt\hbox{$\scriptscriptstyle\smile$}}
\over D}}}{_2^{\rm{T}}}, \cdots ,{\bf{\mathord{\buildrel{\lower3pt\hbox{$\scriptscriptstyle\smile$}}
\over D}}}{_{L}^{\rm{T}}}]^{\rm{T}}}$.
Second, we have ${{\bf{\mathord{\buildrel{\lower3pt\hbox{$\scriptscriptstyle\smile$}}
\over D}}}_{\tilde \Omega}} = {\left[ {{\bf{\mathord{\buildrel{\lower3pt\hbox{$\scriptscriptstyle\smile$}}
\over D}}}_{{\tilde \Omega (1)}}^{\rm{T}},{\bf{\mathord{\buildrel{\lower3pt\hbox{$\scriptscriptstyle\smile$}}
\over D}}}_{{\tilde \Omega (2)}}^{\rm{T}}, \cdots ,{\bf{\mathord{\buildrel{\lower3pt\hbox{$\scriptscriptstyle\smile$}}
\over D}}}_{{\tilde \Omega ({| \tilde \Omega  |_c})}}^{\rm{T}}} \right]^{\rm{T}}}$ and ${{\bf{\Psi }}_{\tilde \Omega}} = {\left[ {{\bf{\Psi }}_{{\tilde \Omega (1)}}^{\rm{}},{\bf{\Psi }}_{{\tilde \Omega (2)}}^{\rm{}}, \cdots ,{\bf{\Psi }}_{{\tilde \Omega ({| \tilde \Omega  |_c})}}^{\rm{}}} \right]^{\rm{}}}$, where ${\tilde \Omega(1)} < {\tilde \Omega (2)} <  \cdots  < {\tilde \Omega ({| \tilde \Omega  |_c})}$ are elements in the set $\tilde \Omega$.
Third, ${\Pi ^s}\left( {\cdot} \right)$ is a set, whose elements are the indices of the largest $s$ elements of its argument. Finally, to reliably acquire the channel sparsity level, we stop the iteration if ${\left\| {{{\bf{R}}^{k }}} \right\|_F} > {\left\| {\bf{R}}_{s-1} \right\|_F}$ or ${\left\| {{{{\bf{\mathord{\buildrel{\lower3pt\hbox{$\scriptscriptstyle\smile$}}
\over D}}}}_{\tilde l}}} \right\|_F} \le \sqrt {MR} {p_{{\rm{th}}}}$, where ${\left\| {{{{\bf{\mathord{\buildrel{\lower3pt\hbox{$\scriptscriptstyle\smile$}}
\over D}}}}_{\tilde l}}} \right\|_F}$ is the smallest ${\left\| {{{{\bf{\mathord{\buildrel{\lower3pt\hbox{$\scriptscriptstyle\smile$}}
\over D}}}}_{ l}}} \right\|_F}$ for ${ l} \in {\tilde \Omega^k}$, and $p_{\rm{th}}$ is the noise floor according to \cite{th}.
The proposed stopping criteria will be further discussed in Section \ref{convergence}. % to find the set $\Gamma$ containing indices $\left\{ l \right\}_{l = 1}^L$ associated with the first $\cal{T}$ largest ${\left\| {{{\bf{A}}_l}} \right\|_F}$.

Here we further explain the main steps in {Algorithm 1} as follows. First, for {\textbf{step 2.1}$ \sim$\textbf{2.7}}, the ASSP algorithm aims to acquire the solution $\bf{D}$ to (9) with the fixed sparsity level $s$ in a greedy way, which is similar to the classical SP algorithm. Second, ${\left\| {{{\bf{R}}^{k - 1}}} \right\|_F} \le {\left\| {\bf{R}}^{k} \right\|_F}$ indicates that the $s$-sparse solution $\bf{D}$ to (9) has been obtained, and then the sparsity level is updated to find the $(s+1)$-sparse solution $\bf{D}$. Finally, if the stopping criteria are met, the iteration quits, and we consider the estimated solution to (\ref{equ:common}) with the last sparsity level as the estimated channels, i.e., $\widehat {\bf{D}}  = {{\bf{\mathord{\buildrel{\lower3pt\hbox{$\scriptscriptstyle\smile$}}
\over D}}}_{s-1}}$.

%\emph{Step 17$ \sim$21} and \emph{step 32} give the stopping conditions, and the detail physical meaning will be provided in Section \ref{convergence}. When iteration steps into \emph{Step 22$ \sim$26}, the new stage is started with the updated sparsity level. It is worth noting that since \emph{step 10$ \sim$16} and \emph{step 28$ \sim$30} works in the greedy way provided the fixed sparsity level (for each stage, the used sparsity level $\cal{T}$ is fixed), the residue decrease with the number of iteration increase, and ${\left\| {{{\bf{R}}^{i - 1}}} \right\|_F} \le {\left\| {\bf{R}} \right\|_F}$ indicates the iteration converges with the current sparsity level.

Compared to the SP algorithm and the model-based SP algorithm \cite{model}, the proposed ASSP algorithm has the following distinctive features:
\begin{itemize}
  \item The classical SP algorithm reconstructs one high-dimensional sparse vector from one low-dimensional measurement vector without exploiting the structured sparsity of the sparse vector. The model-based SP algorithm reconstructs one high-dimensional sparse vector from one low-dimensional measurement vector by exploiting the structured sparsity of the sparse vector for improved performance. In contrast, the proposed ASSP algorithm recovers the high-dimensional sparse matrix with the inherently structured sparsity from the low-dimensional measurement matrix, whereby the inherently structured sparsity of the sparse matrix is exploited for the improved matrix reconstruction performance.
  \item Both the classical SP algorithm and model-based SP algorithm require the sparsity level as the priori information for reliable sparse signal reconstruction. In contrast, the proposed ASSP algorithm does not need this priori information, since it can adaptively acquire the sparsity level of the structured sparse matrix. By exploiting the practical physical property of wireless channels, the proposed stopping criteria enable ASSP algorithm to estimate channels with good mean square error (MSE) performance, which will be detailed in Section \ref{convergence}. Moreover, simulation results in Section \ref{section_sim} verify its accurate acquisition of channel sparsity~level.
\end{itemize}
Hence, the conventional SP algorithm and model-based SP algorithm can be considered as two special cases of the proposed ASSP algorithm.

It should be pointed out that, most of the state-of-the-art CS-based channel estimation schemes usually require the channel sparsity level as the priori information for reliable channel estimation~\cite{{CS_dai},{BMSB},{GuiG}}. In contrast, the proposed ASSP algorithm removes this unrealistic assumption, since it can adaptively acquire the sparsity level of wireless MIMO channels.

\vspace*{-3.0mm}
\subsection{Space-Time Adaptive Pilot Scheme}\label{D}
\begin{figure}[!tp]
     \centering
     %\vspace*{-10.0mm}
     \includegraphics[width=9.3cm, keepaspectratio]
     {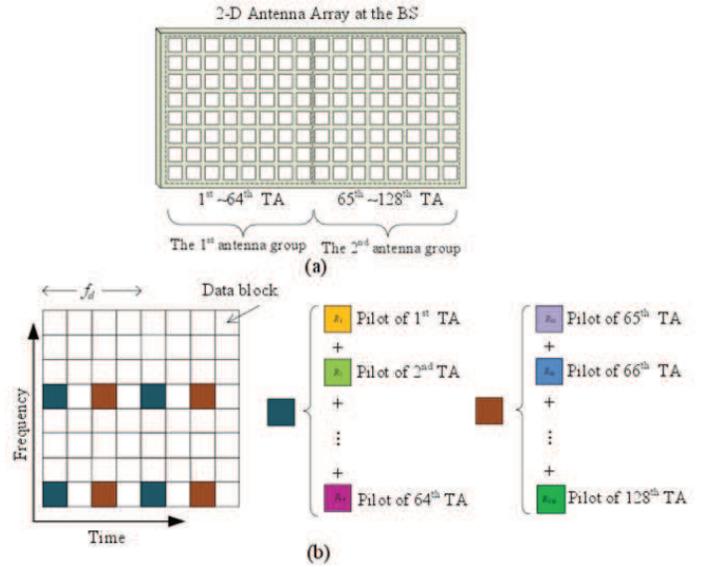}
      \vspace*{-3.0mm}
    \caption{Space-time adaptive pilot scheme, where $M=128$, $N_G=2$, $f_d=4$, and the adjacent antenna spacing $\lambda/2 $ are considered as an example. (a) 2-D antenna array at the BS; (b) Space-time adaptive pilot scheme.}%; (b): Proposed superimposed pilot schemewhere horizontal axis is path delays, and vertical %axis is delay gains
     \label{fig:hybrid}
      \vspace*{-2.0mm}
\end{figure}

As we have demonstrated in Section \ref{SCS}, the spatial common sparsity of MIMO channels is due to the co-located antenna array at the BS. However, for massive MIMO with large antenna array, such common sparsity may not be ensured for antennas spaced apart. To address this problem,
%such common path delays  property is not limited to the linear unform array \cite{scaleMIMO}.
%To this end, we propose the space-time division based superimposed pilot strategy. In particular,
we propose that $M$ transmit antennas are divided into $N_G$ antenna groups, where $M_G=M/N_G$ antennas with close distance in the spatial domain are assigned to the same antenna group, so that the spatial common sparsity of wireless MIMO channels in each antenna group can be guaranteed. For example, we consider the $M = 128$ planar antenna array as shown in Fig. \ref{fig:hybrid} (a), which can be divided into two array groups according to the criterion above. If we consider $f_c=2~\rm{GHz}$, $f_s=10~\rm{MHz}$, and the maximum distance for a pair of antennas in each antenna group as shown in Fig. \ref{fig:hybrid} (a) is $4\sqrt{2}\lambda$, their maximum difference of path delays from the common scatterer is $\frac{{{4\sqrt{2}\lambda}}}{{{c}}} = 4\sqrt{2}/f_c=0.0028~\mu s$, which is negligible compared with the system sample period $T_s=1/f_s=0.1~\mu s$. For a certain antenna group, pilots of different transmit antennas are non-orthogonal and occupy the identical subcarriers, while pilots of different antenna groups are orthogonal in the time domain or frequency domain, which can be illustrated in Fig. \ref{fig:hybrid} (b). For the specific parameter $N_G$, we should consider the geometry and scale of the antenna array at the BS, $f_c$, and $f_s$.
%In this way, pilots of all $M$ transmit antennas, which are transmitted in one time slot previously, are dispersed to $N_G$ time slots currently. Correspondingly, the ratio of pilot overhead reduces and the spectral efficiency boosts.

On the other hand, wireless MIMO channels exhibit the temporal correlation. Such temporal channel correlation indicates that during the coherence time of path gains, channels in several successive OFDM symbols can be considered to be quasi-static, and the channel estimation in one OFDM symbol can be used to estimate channels of several adjacent OFDM symbols. This motivates us to further reduce the pilot overhead and increase the available spectrum and energy resources for effective data transmission. To be specific, as illustrated in Fig. \ref{fig:hybrid}, every $f_d$ adjacent OFDM symbols share the common pilots, where $f_d$ is determined by the coherence time of path gains or the mobility of served~users.

By exploiting such temporal channel correlation, we can use large $f_d$ to reduce the pilot overhead. To estimate channels of OFDM symbols without pilots, we can use interpolation algorithms according to the estimated channels of adjacent OFDM symbols with pilots, e.g., we can adopt the linear interpolation algorithm as follows
      %\vspace*{-2.0mm}
\begin{equation}\label{equ:interp} %for OFDM symbols without pilot signals, acquire the corresponding CSI by
{\hat {\bf{h}}}_{m,r}=[(f_p+1-r){{\hat {\bf{h}}}}_{m,1}+(r-1){{\hat {\bf{h}}}}_{m,f_p+1}]/f_p,
%\vspace*{-2.0mm}
\end{equation}
where $1<r \le f_p$, ${{\hat {\bf{h}}}}_{m,1}$ and ${{\hat {\bf{h}}}}_{m,f_p+1}$ are the estimated channels of the first and ($f_p+1$)th OFDM symbols, respectively, and ${\hat {\bf{h}}}_{m,r}$ is the interpolated channel estimation of the $r$th OFDM symbol. % CSI by using the linear interpolation algorithm.
%For typical massive MIMO systems, the mobility of supported users can be low.

The proposed space-time adaptive pilot scheme considers both the geometry of the antenna array at the BS and the mobility of served users, which can achieve the reliable channel estimation and further reduce the required pilot overhead. For the space-time adaptive pilot scheme, the proposed ASSP algorithm is used at the user to estimate channels associated with different transmit antennas in each antenna group, where the received pilots associated with different antenna groups are processed separately. In Section \ref{section_sim}, the simulation results will show that the proposed space-time adaptive pilot scheme can further reduce the required pilot overhead with a negligible performance loss, even for the high speed scenario where the users' mobile velocity is 60 km/h.%Besides, we also introduce the pilot insertion frequency $f_d$ as shown in Fig. \ref{fig:hybrid}, which implies that pilot sequences from the same antenna group are inserted every $f_d$ OFDM symbols. In practical massive MIMO systems, different adopted $f_d$'s are adaptive to various scenarios.

%Finally, the proposed space-time division based superimposed pilot strategy should be addressed below. The motivation of pilots dispersed in multiple time slots in the proposed strategy is essentially different from that in massive MIMO. Recall that the number of time slots devoted to pilots in conventional massive MIMO must be at least as $M/N_\text{Coh}$ due to the limited coherence frequency $N_\text{Coh}$ and the large number of $M$. Hence the frequency-domain orthogonal pilots based massive MIMO systems are confined to slow time-varying channels. By contrast, the goal that we develop the space-time division based superimposed pilot strategy is to improve the spectral efficiency by using the strong correlated in slow time-varying channels. It should be pointed out that the number of time slots devoted to pilots $N_G$ in the proposed space-time division based superimposed pilot strategy is flexible according to the mobility of the supported users.
%\vspace*{-3.0mm}

\subsection{Channel Estimation in Multi-Cell Massive MIMO}\label{multi-cell}
In this subsection, we extend the proposed channel estimation scheme from the single-cell scenario to the multi-cell scenario. We consider a cellular network composed of ${\cal{L}}=7$ hexagonal cells, each consisting of a central $M$-antenna BS and $K$ single-antenna users that share the same bandwidth, where the users of the central target cell suffer from the interference of the surrounding ${\cal{L}}-1$ interfering cells. One straightforward solution to solve the pilot contamination from the interfering cells is the frequency-division multiplexing (FDM) scheme, i.e., pilots of adjacent cells are orthogonal in the frequency domain. FDM scheme can perfectly mitigate the pilot contamination if the training time used for channel estimation is less than the channel coherence time, but it can lead to the ${\cal{L}}$ times pilot overhead in multi-cell system than that in single-cell system. An alternative solution is the time-division multiplexing (TDM) scheme \cite{multicell}, where pilots of adjacent cells are transmitted in different time slots. The pilot overhead with TDM scheme in multi-cell scenario is the same with that in single-cell scenario. However, the downlink precoded data from adjacent cells may degrade the channel estimation performance of users in the target cell. In Section \ref{section_sim}, we will verify that the TDM scheme can be the viable approach to mitigate the pilot contamination in multi-cell FDD massive MIMO systems due to the obviously reduced pilot overhead and the slightly performance loss compared to the FDM scheme.

% ֮ǰÎÒÃÇÌÖÂ۵ĵ¥Ð¡Çømassive mimo,ÕâÀïÎÒÃÇÀ©Õ¹µ¥Ð¡Çøµ½¶àСÇø³¡¾°¡£ÎÒÃÇ¿¼ÂÇLСÇø¡£µÚÒ»ÖÖ£¬ÎÒÃÇ×îÓŵķ½·¨£¬ÎÒÃdzÆ֮ΪFDM Ò²¾ÍÊÇ¡£¡£¡£¡£Õý½»
% µÚ¶þÖÖ£¬ÎÒÃDzÉÓÃÑÓʱÑо¿ÊÒ¡£¡£¡£¡£TDM¡£ µÚÈýÖÖ NCDM
%\vspace*{-3.0mm}
\section{Performance Analysis}\label{performance}
In this section, we first provide the design of the proposed non-orthogonal pilot scheme for reliable channel estimation under the framework of CS theory. Then we analyze the convergence analysis and complexity of the proposed ASSP algorithm. % Finally, the spectrum-energy efficiency improvement by the proposed channel estimation scheme is discussed.
%\vspace*{-2.0mm}
\subsection{Non-Orthogonal Pilot Design Under the Framework of CS Theory}\label{pilot_design}
In CS theory, design of the sensing matrix ${\bf{\Psi }}$ in (\ref{equ:common}) is very important to effectively and reliably compress the high-dimensional sparse signal $\bf{D}$. For the problem of channel estimation, the design of ${\bf{\Psi }}$ is converted to the design of the pilot placement $\xi$ and the pilot sequences $\left\{ {{{\bf{p}}_m}} \right\}_{m = 1}^M$, since the sensing matrix ${\bf{\Psi }}$ is only determined by the parameters $\xi$ and $\left\{ {{{\bf{p}}_m}} \right\}_{m = 1}^M$.
According to CS theory, the small column correlation of ${\bf{\Psi }}$ is desired for the reliable sparse signal recovery~\cite{STR_CS}, which enlightens us to appropriately design $\xi$ and $\left\{ {{{\bf{p}}_m}} \right\}_{m = 1}^M$.

%To guarantee the reliable sparse signal recovery, the sensing matrix ${\bf{\Psi }}$ should satisfy
%\begin{equation}\label{equ:coherence}
%P < \frac{1}{2}\left( {1 + \frac{1}{{\mu \left( {\bf{\Psi }} \right)}}} \right),
%\end{equation}
%according to \cite{STR_CS}, where $P$ denotes the sparsity level, and $\mu ({\bf{\Psi }}) $ denotes the maximal column correlation of ${\bf{\Psi }}$, i.e.,
%\begin{equation}\label{equ:coherence2}
%\begin{small}
%\mu ({\bf{\Psi }}) = \mathop {{\rm{max}}}\limits_{m_1 \ne m_2~ {\rm{or}} ~l_1\ne l_2} \frac{{\left| ({{\bm{\psi }}_{{m_1},{l_1}})^{\rm{H}}{{\bm{\psi }}_{{m_2},{l_2}}}} %\right|}}{{{{\left\| {{\bm{\psi }}_{{m_1},{l_1}}} \right\|}_2}{{\left\| {{{\bm{\psi }}_{{m_2},{l_2}}}^{}} \right\|}_2}}},
%\end{small}
%\end{equation}
%where $1 \le {m_1} , {m_2} \le M$ and $1 \le {l_1} , {l_2} \le L$.

%From (\ref{equ:block_guarantee}), we find that lower column correlation
For the specific pilot design, we commence by considering the design of $\left\{ {{{\bf{p}}_m}} \right\}_{m = 1}^M$ to achieve the small cross-correlation for columns of ${{\bf{\Psi }}_l}$ given any $l$, since this kind of cross-correlation is only determined by $\left\{ {{{\bf{p}}_m}} \right\}_{m = 1}^M$,~i.e.,
\begin{equation}\label{equ:proof}
\begin{array}{l}
\!\!\!\!\!\!\!\!({\bm{\psi}}_{{m_1},l})^{\rm{H}}{{\bm{\psi}}_{{m_2},l}}= ({\bf{\Psi }}_{{l}}^{(m_1)})^{\rm{H}}{\bf{\Psi }}_{{l}}^{(m_2)} = ({\bf{\Phi }}_{{m_1}}^{(l)})^{\rm{H}}{\bf{\Phi }}_{{m_2}}^{(l)}\\
\!\!\!\!\!\!\!\! {\kern 59pt}  = {({{\bf{p}}_{{m_1}}} \circ {\bf{F}}_p^{(l)})^{\rm{H}}}({{\bf{p}}_{{m_2}}} \circ {\bf{F}}_p^{(l)})   = ({\bf{p}}_{{m_1}})^{\rm{H}}{{\bf{p}}_{{m_2}}}.
\end{array}
\end{equation}
where ${\bf{F}}_p={\left. {{{\bf{F}}_L}} \right|_\xi }$ and $1 \le m_1 < m_2 \le M$.

To achieve the small $\left| ({{\bm{\psi}}_{{m_1},l})^{\rm{H}}{{\bm{\psi}}_{{m_2},l}}} \right|$, we consider $\left\{ {{\theta _{\kappa ,m}}} \right\}_{\kappa  = 1,m = 1}^{{N_p},M}$ to follow the independent and identically distributed (i.i.d.) uniform distribution ${\cal{U}}\left[ {0,2\pi } \right)$, where ${e^{j{\theta}_{\kappa,m} }}$ denotes the $\kappa$th element of $ {{{\bf{p}}_m}}\in \mathbb{C}^{{N_p} \times 1}$. For the proposed pilot sequences, the $l_2$-norm of each column of ${\bf{\Psi }}$ is a constant, i.e., ${\left\| {{{\bm{\psi }}_{m,l}}} \right\|_2} = \sqrt {{N_p}} $. Meanwhile, we have
\begin{equation}\label{equ:coherence2}
\mathop {\lim }\limits_{{N_p} \to \infty } \frac{{\left|( {{\bm{\psi }}_{{m_1},{l}})^{\rm{H}}{{\bm{\psi }}_{{m_2},{l}}}} \right|}}{{{{\left\| {{{\bm{\psi }}_{{m_1},{l}}}} \right\|}_2}{{\left\| {{{\bm{\psi }}_{{m_2},{l}}}^{}} \right\|}_2}}} = \mathop {\lim }\limits_{{N_p} \to \infty } \frac{{({\bf{p}}_{{m_1}})^{\rm{H}}{{\bf{p}}_{{m_2}}}}}{{{N_p}}} = 0,
\end{equation}
which indicates that for the limited $N_p$ in practice, the proposed pilot sequences can achieve the good cross-correlation of columns of ${{\bf{\Psi }}_l}$ for any $l$ according to the random matrix theory (RMT).
%Thus $\upsilon \left( {\bf{\Psi }} \right)=0$.
%\end{proof}

Given the proposed $\left\{ {{{\bf{p}}_m}} \right\}_{m = 1}^M$, we further investigate the cross-correlation of ${\bm{\psi }}_{{m_1},{l_1}}$ and ${{\bm{\psi }}_{{m_2},{l_2}}}$ with $l_1 \ne l_2$, which enlightens us to design $\xi$ to achieve the small $\left|({\bm{\psi }}_{{m_1},{l_1}})^{\rm{H}}{{\bm{\psi }}_{{m_2},{l_2}}}\right| $. % ($1\le m_1 \le M$, $1\le m_2 \le M$, and $1 \le l_1 < l_2 \le L$).
In typical massive MIMO systems (e.g., $M\ge 64$), we usually have $N_p>L$, which is due to the two following reasons. First, since the number of pilots for estimating the channel associated with one transmit antenna is at least one, the number of the total pilot overhead $N_p$ can be at least $64$. Second, since the maximum channel delay spread is $3 \sim 5~\mu s$ and the typical system bandwidth is $10$ MHz if we refer to the LTE-Advanced system parameters, we have $L\le 64$ \cite{3GPP}. Based on the condition of $N_p>L$, we propose to adopt the widely used uniformly-spaced pilots with the pilot interval $\left\lfloor {\frac{N}{{{N_p}}}} \right\rfloor$ to acquire the small $\left|({\bm{\psi }}_{{m_1},{l_1}})^{\rm{H}}{{\bm{\psi }}_{{m_2},{l_2}}}\right| $.
Specifically, we consider $\xi $ is selected from the set of $\left\{ {1,2, \cdots ,N} \right\}$ with the equal interval, and the inner product of ${\bm{\psi }}_{{m_1},{l_1}}$ and ${{\bm{\psi }}_{{m_2},{l_2}}}$ can be expressed as
\begin{equation}\label{equ:proof2}
\!\!\!\!\!\!\!\!\begin{array}{l}
({\bm{\psi }}_{{m_1},{l_1}})^{\rm{H}}{{\bm{\psi }}_{{m_2},{l_2}}} = ({\bf{\Phi }}_{{m_1}}^{({l_1})})^{\rm{H}}{\bf{\Phi }}_{{m_2}}^{({l_2})} = {({{\bf{p}}_{{m_1}}} \circ {\bf{F}}_p^{({l_1})})^{\rm{H}}}({{\bf{p}}_{{m_2}}} \circ {\bf{F}}_p^{({l_2})})\\
 {\kern 74pt} = \sum\limits_{\kappa  = 1}^{{N_p}} {\exp {{\left( {j\frac{{2\pi }}{N}{l_1}I\left( \kappa  \right) + j{\theta _{\kappa ,m_1}}} \right)}^{\rm{H}}}} \\ {\kern 90pt}
 \times {\exp {{\left( {j\frac{{2\pi }}{N}{l_2}I\left( \kappa  \right) + j{\theta _{\kappa ,m_2}}} \right)}}} \\
 {\kern 74pt}  = \sum\limits_{\kappa  = 1}^{{N_p}} {\exp \left( {j\frac{{2\pi }}{N}\tilde lI\left( \kappa  \right)}+ {j\Delta {\theta _{\kappa ,m}}} \right)},
\end{array}
\end{equation}
where $\left\{ {I\left( \kappa  \right)} \right\}_{\kappa  = 1}^{{N_p}} = \xi $ is the indices set of pilot subcarriers, $1 \le {\tilde l}=l_2 -l_1 \le L-1$, and $\Delta {\theta _{\kappa ,m}}{\rm{ = }}{\theta _{\kappa ,{m_2}}}{\rm{ - }}{\theta _{\kappa ,{m_1}}}$.
Furthermore, since $\left\{ {I\left( \kappa  \right)} \right\}_{\kappa  = 1}^{{N_p}}$ is selected from the set of $\left\{ {1,2, \cdots ,N} \right\}$ with the equal interval $\left\lfloor {\frac{N}{{{N_p}}}} \right\rfloor$, $I\left( \kappa  \right) = {I_0} +( \kappa-1)\left\lfloor {\frac{N}{{{N_p}}}} \right\rfloor  $ for $1\le \kappa \le N_p$, where $I_0$ is the subcarrier index of the first pilot with $1 \le {I_0} < \left\lfloor {\frac{N}{{{N_p}}}} \right\rfloor $. Hence, (\ref{equ:proof2}) can be also expressed as
\vspace*{-2.0mm}
\begin{equation}\label{equ:proof3}
\!\!\!\begin{array}{l}
({\bm{\psi }}_{{m_1},{l_1}})^{\rm{H}}{{\bm{\psi }}_{{m_2},{l_2}}} = \\ {\kern 15pt}
 {\sum\limits_{\kappa  = 1}^{{N_p}} {\exp \left( {j\frac{{2\pi }}{N}\tilde l\left( {{I_0} + (\kappa  - 1)\left\lfloor {\frac{N}{{{N_p}}}} \right\rfloor } \right) + j\Delta {\theta _{\kappa ,m}}} \right)} }.
\end{array}
\end{equation}
Let $\varepsilon  = \frac{N}{{{N_p}}} - \left\lfloor {\frac{N}{{{N_p}}}} \right\rfloor $ with $\varepsilon  \in \left[ {0,1} \right)$, we can further obtain
\vspace*{-2.0mm}
\begin{equation}\label{equ:proof4}
\!\!\!\!\!\!\!\begin{array}{l}
({\bm{\psi }}_{{m_1},{l_1}})^{\rm{H}}{{\bm{\psi }}_{{m_2},{l_2}}} = {c_0}\sum\limits_{\kappa  = 1}^{{N_p}} {\exp \left( {j\frac{{2\pi }}{N}\tilde l\kappa \left( {\frac{N}{{{N_p}}} - \varepsilon } \right) + {j\Delta {\theta _{\kappa ,m}}}} \right)},% = {C_1}{C_2}\sum\limits_{\kappa  = 1}^{{N_p}} {\exp \left( {j\frac{{2\pi }}{{{N_p}}}\tilde l\kappa \left( {1 - \frac{{{N_p}}}{N}\varepsilon  + \left( {{N_p} - 1} \right)r\Delta } \right)} \right)}.
\end{array}
\end{equation}
where $c_0=\exp \left( {j\frac{{2\pi }}{N}\tilde l\left( {{I_0} - \left\lfloor {\frac{N}{{{N_p}}}} \right\rfloor } \right)} \right)$.
To investigate $\left|({\bm{\psi }}_{{m_1},{l_1}})^{\rm{H}}{{\bm{\psi }}_{{m_2},{l_2}}}\right| $ with $\l_1 \ne l_2$, we consider the following two cases.
For the first case, if $m_1=m_2$, then $\Delta\theta _{\kappa ,m} = 0$, and (\ref{equ:proof4}) can be simplified as
\vspace*{-2.0mm}
 \begin{equation}\label{equ:proof5}
\begin{array}{l}
({\bm{\psi }}_{{m_1},{l_1}})^{\rm{H}}{{\bm{\psi }}_{{m_2},{l_2}}}  = {c_0}\sum\limits_{\kappa  = 1}^{{N_p}} {\exp \left( {j\frac{{2\pi }}{{{N_p}}}\tilde l\kappa \left( {1 - \eta \varepsilon } \right)} \right)},
\end{array}
\end{equation}
where $\eta  = \frac{{{N_p}}}{N}<1$ denotes the pilot occupation ratio. Thus, ${\eta \varepsilon  \approx 0}$, and we can obtain
 \begin{equation}\label{equ:proof6}
\mathop {\lim }\limits_{{N_p} \to \infty } \frac{({{\bm{\psi }}_{{m_1},{l_1}})^{\rm{H}}{{\bm{\psi }}_{{m_2},{l_2}}}}}{{{N_p}}} = \mathop {\lim }\limits_{{N_p} \to \infty } \frac{{c_0\left( {1 - {e^{j2\pi \tilde l\left( {1 - \eta \varepsilon } \right)}}} \right)}}{{{N_p}\left( {1 - {e^{j\frac{{2\pi }}{{{N_p}}}\tilde l\left( {1 - \eta \varepsilon } \right)}}} \right)}} = 0,
\end{equation}
where ${e^{j\frac{{2\pi }}{{{N_p}}}\tilde l\left( {1 - \eta \varepsilon } \right)}} \approx e^{\left( {j\frac{{2\pi }}{{{N_p}}}\tilde l} \right) }\ne 1$ guarantees the validity of (\ref{equ:proof6}) due to $1 \le {\tilde l} \le L-1$ and $L<N_p$.
For the second case, if $m_1 \ne m_2$, then (\ref{equ:proof4}) can be expressed~as
\begin{equation}\label{equ:proof7}
\begin{array}{l}
({\bm{\psi }}_{{m_1},{l_1}})^{\rm{H}}{{\bm{\psi }}_{{m_2},{l_2}}}= \sum\limits_{\kappa  = 1}^{{N_p}} {\exp \left( {j{\tilde \theta _\kappa }} \right)},
\end{array}
\end{equation}
where ${\tilde \theta _\kappa }=\frac{{2\pi }}{N}\tilde lI\left( \kappa  \right){\rm{ + }}\Delta {\theta _{\kappa ,m}}$ for $1\le \kappa \le N_p$ follow the i.i.d. distribution ${\cal{U}}\left[ {0,2\pi } \right)$. Similar to (\ref{equ:coherence2}), we further have
\vspace*{-3.0mm}
\begin{equation}\label{equ:proof8}
\begin{array}{l}
\mathop {\lim }\limits_{{N_p} \to \infty } \frac{({{\bm{\psi }}_{{m_1},{l_1}})^{\rm{H}}{{\bm{\psi }}_{{m_2},{l_2}}}}}{{{N_p}}} = \mathop {\lim }\limits_{{N_p} \to \infty } \frac{{\sum\limits_{\kappa  = 1}^{{N_p}} {\exp \left( {j{{\tilde \theta }_\kappa }} \right)} }}{{{N_p}}} = 0.
\end{array}
\end{equation}

According to RMT, the asymptotic orthogonality of (\ref{equ:coherence2}), (\ref{equ:proof6}), and (\ref{equ:proof8}) indicates that the proposed $\xi$ and $\left\{ {{{\bf{p}}_m}} \right\}_{m = 1}^M$ can achieve the good cross-correlation between any two columns of ${{\bf{\Psi }}}$ with the limited $N_p$ in practice. Moreover, compared with the conventional random pilot placement scheme widely used in CS-based channel estimation schemes \cite{CS_Sparse}, the proposed uniformly-spaced pilot placement scheme can be more easily implemented in practical systems due to its regular pattern. Moreover, it can also facilitate massive MIMO to be backward compatible with current cellular networks, since the uniformly-spaced pilot placement scheme has been widely used in existing cellular networks \cite{LTE_review}. Finally, its reliable sparse signal recovery performance can be verified through simulations in Section~\ref{section_sim}.

%for any $l$ according to the random matrix theory (MRT).

%To sum up, the proposed pilot scheme could enable the small column correlation of the sensing matrix ${\bf{\Phi }}$,
 %\vspace*{-3.0mm}
\subsection{Convergence Analysis of Proposed ASSP Algorithm}\label{convergence}
For the proposed ASSP algorithm, we first provide the convergence with the correct sparsity level $s = P$. Then we provide the convergence for the case of $s \ne P$, where the proposed stopping criteria are also discussed. It should be pointed out that conventional SP algorithm and model-based SP algorithm analyze the convergence for the recovery of a single sparse vector. By contrast, we provide the convergence for the reconstruction of structured sparse matrix.

The convergence for the case of $s= P$ can be guaranteed due to the following theorem.

\begin{theorem}\label{T1}
 For ${\bf{Y}} = {\bf{\Psi D}}+{\bf{W}}$ and the ASSP algorithm with the sparsity level $s= P$, we have
 \begin{equation}\label{equ:app001}
 \begin{small}
{\left\| {{\bf{D}}{\rm{ - }}{\bf{\hat D}}} \right\|_F} \le {c}_P{\left\| {\bf{W}} \right\|_F},
\end{small}
\end{equation}
 \begin{equation}\label{equ:app002}
 \begin{small}
{\left\| {{{\bf{R}}^k}} \right\|_F} < {{c'}_P}{\left\| {{{\bf{R}}^{k - 1}}} \right\|_F} +  {{c''}_P}{\left\| {\bf{W}} \right\|_F},
\end{small}
\end{equation}
where ${\bf{\hat D}}$ is the estimation of $\bf{D}$ with $s=P$, and ${c}_P$, ${{c'}_P}$, and ${{c''}_P}$ are constants.
\end{theorem}
Here ${c}_P$, ${{c'}_P}$, and ${{c''}_P}$ are determined by the structured restricted isometry property (SRIP) constants $\delta_P $, $\delta_{2P} $, and $\delta_{3P} $, which will be further detailed in Appendix~\ref{TPa}. The proof of Theorem~\ref{T1} will be provided in Appendix~\ref{TPa}.

Moreover, we investigate the convergence of the case with $s\ne P$. We consider ${\bf{D}} = {\left. {\bf{D}} \right\rangle _s} + \left( {{\bf{D}} - {{\left. {\bf{D}} \right\rangle }_s}} \right)$, where the matrix ${\left. {\bf{D}} \right\rangle _s}$ preserves the largest $s$ sub-matrices $\left\{ {{{\bf{D}}_l}} \right\}_{l = 1}^L$ according to their $F$-norms and sets other sub-matrices to $\bf{0}$. In this way, (\ref{equ:common}) can be further expressed as
 \begin{equation}\label{equ:convgence4}
{\bf{Y}} = {\bf{\Psi }}{\left. {\bf{D}} \right\rangle _s} + {\bf{\Psi }}\left( {{\bf{D}} - {{\left. {\bf{D}} \right\rangle }_s}} \right) + {\bf{W}} = {\bf{\Psi }}{\left. {\bf{D}} \right\rangle _s} + {\bf{W'}},
\end{equation}
where ${\bf{W'}} =  {\bf{\Psi }}\left( {{\bf{D}} - {{\left. {\bf{D}} \right\rangle }_s}} \right) + {\bf{W}}$.
For the case of $s\ne P$, we may not reliably reconstruct the $P$-sparse signal $\bf{D}$ even the $s$-sparse signal ${\bf{\mathord{\buildrel{\lower3pt\hbox{$\scriptscriptstyle\smile$}}
\over D}}}_s$ is estimated. However, with the appropriate SRIP, Theorem~\ref{T1} indicates that we can acquire partial correct support set from the estimated $s$-sparse matrix, i.e., ${\Omega _s} \cap {\Omega _T} \ne \phi $, where ${\Omega _s}$ is the support set of the estimated $s$-sparse matrix, ${\Omega _T}$ is the true support set of ${\bf{ D}}$, and $\phi$ denotes the null set. Hence ${\Omega _s} \cap {\Omega _T} \ne \phi $ can reduce the number of iterations for the convergence with the sparsity level $s+1$, since the first iteration with the sparsity level $s+1$ uses ${\Omega _s}$ as the priori information (\textbf{Step 2.2} in Algorithm 1). It should be pointed out that the proof of Theorem~\ref{T1} does not rely on the estimated support set with the last sparsity level.% Even for the worst case that ${\Omega _{P-1}} \cap {\Omega _T} = \phi $, the proposed ASSP algorithm can still reliably acquire the $P$-sparse signal, since the proof of Theorem~\ref{T1} does not rely on ${\Omega _{P-1}}$.

Additionally, by exploiting the practical channel property, the proposed stopping criteria enable ASSP algorithm to achieve good MSE performance, and we will discuss the proposed stopping criteria as follows.
The stopping criterion ${\left\| {{{\bf{R}}^{k }}} \right\|_F} > {\left\| {\bf{R}}_{s-1} \right\|_F}$ is clear as it implies that the residue of the current sparsity level is larger than that of the last sparsity level, and stopping the iteration can help the algorithm to acquire the good MSE performance. On the other hand, the stopping criterion ${\left\| {{{{\bf{\mathord{\buildrel{\lower3pt\hbox{$\scriptscriptstyle\smile$}}
\over D} }}}_{\tilde l}}} \right\|_F} \le \sqrt {M_GR} {p_{{\rm{th}}}}$ implies that the $\tilde l$th path is dominated by AWGN. That is to say, the channel sparsity level is over estimated, although MSE performance with the current sparsity level is better than that with the last sparsity level. Actually, the improvement of MSE performance is due to ``reconstructing" noise.
\subsection{Computational Complexity of ASSP Algorithm}\label{complexity}
\begin{figure}[!tp]
     \centering
     % \vspace*{-10.0mm}
     \includegraphics[width=9cm, keepaspectratio]
         {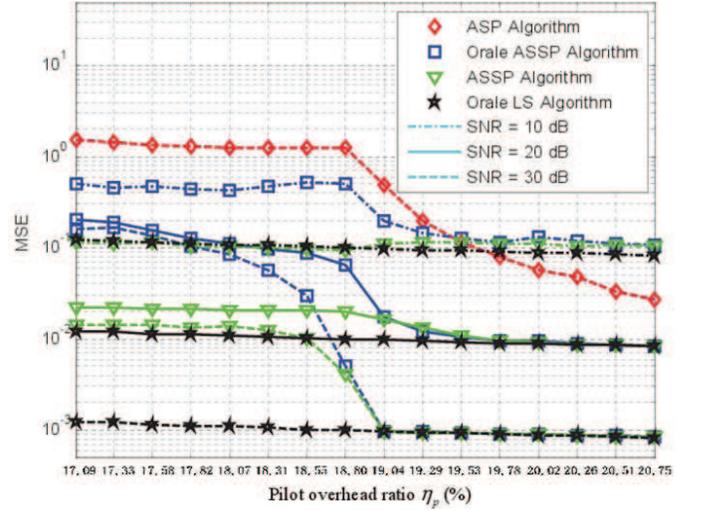}
     \vspace*{-3.0mm}
    \caption{MSE performance comparison of different channel estimation algorithms against pilot overhead ratio and SNR.}%; (b): Proposed superimposed pilot schemewhere horizontal axis is path delays, and vertical %axis is delay gains
     \label{fig:MSE64_1}
      \vspace*{-2.0mm}
\end{figure}

In each iteration of the proposed ASSP algorithm, the computational complexity mainly comes from the several operations as follows, where the space-time adaptive pilot scheme with $M_G$ transmit antennas in each antenna group is considered. For \textbf{Step 2.1}, the correlation operation has the complexity of ${\cal{O}}(RLM_GN_p)$.
For \textbf{Step 2.2}, both the support merger and ${\Pi ^s}\left( {\cdot} \right)$  have the complexity of ${\cal{O}}(L)$ \cite{algorithm,gao_x}, while the norm operation has the complexity of ${\cal{O}}(RLM_G)$. For \textbf{Step 2.3}, the Moore-Penrose matrix inversion operation has the complexity of ${\cal{O}}(2N_p({{M_Gs}})^2+ ({{M_Gs}})^3 )$ \cite{LS}, ${\Pi ^s}\left( {\cdot} \right)$ has the complexity of ${\cal{O}}(L)$, and the norm operation has the complexity of ${\cal{O}}(RLM_G)$.
For \textbf{Step 2.4}, the Moore-Penrose matrix inversion operation has the complexity of ${\cal{O}}(2N_p({{M_Gs}})^2+ ({{M_Gs}})^3 )$. For \textbf{Step 2.5}, the residue update has the complexity of ${\cal{O}}(RLM_GN_p )$.
To quantitatively compare the computational complexity of different operations, we consider the parameters used in Fig. \ref{fig:MSE64_1} when the performance of the proposed ASSP algorithm approaches that of the oracle LS algorithm. In this case, the ratios of the complexity of the correlation operation, the support merger or ${\Pi ^s}\left( {\cdot} \right)$ operation, the norm operation, and the residue update to that of the Moore-Penrose matrix inversion operation are $2.3\times 10^{-2}$, $1.7\times 10^{-6}$, $5.7\times 10^{-5}$, and $2.3\times 10^{-2}$, respectively.
Therefore, the main computational complexity of the ASSP algorithm comes from the Moore-Penrose matrix inversion operation with the complexity of ${\cal{O}}(2N_p({{M_Gs}})^2+ ({{M_Gs}})^3 )$.

%It should be pointed out that Gauss-Seidel or successive overrelaxation (SOR) method can be used to reduce the complexity of LS solution by one order of magnitude \cite{LS}.

%\subsection{Spectrum-Energy Efficiency}
%\vspace*{-3.0mm}
\section{Simulation Results}\label{section_sim}

\begin{figure*}[tp]
     \centering
      %\vspace*{-10.0mm}
     \includegraphics[width=14cm, keepaspectratio]
         {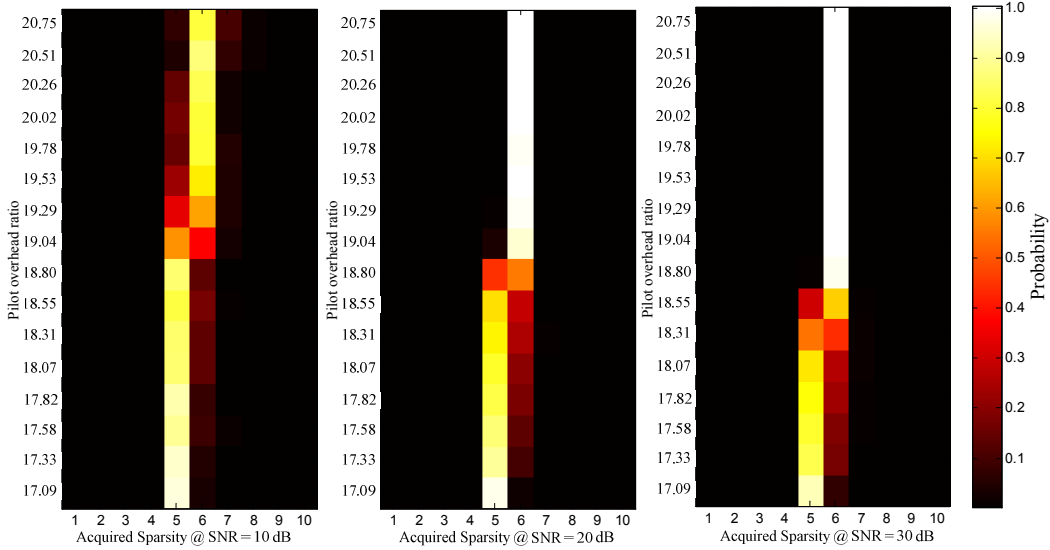}
       %   \vspace*{-3.0mm}
    \caption{Estimated channel sparsity level of the proposed ASSP algorithm against SNR and pilot overhead ratio.}%; (b): Proposed superimposed pilot schemewhere horizontal axis is path delays, and vertical %axis is delay gains
     \label{fig:iter_64}
      %\vspace*{-2.0mm}
\end{figure*}

%the measurement vector of size ${N_p} \approx 2KM$ cannot faithfully recover a ($KM$)-sparse signal of size $ML$ with $K \ll L$
In this section, a simulation study was carried out to investigate the performance of the proposed channel estimation scheme for FDD massive MIMO systems.
To provide a benchmark for performance comparison, we consider the oracle LS algorithm by assuming the true channel support set known at the user and the oracle ASSP algorithm\footnote{The oracle ASSP algorithm is a special case of the proposed ASSP algorithm, where the initial channel sparsity level $s$ is set to the true channel sparsity level, \textbf{Step 2.8} is not performed, the stopping criterion is ${\left\| {{{\bf{R}}^{k - 1}}} \right\|_F}\le {\left\| {\bf{R}}^{k} \right\|_F}$, and $\widehat {\bf{D}}  = {{\bf{\mathord{\buildrel{\lower3pt\hbox{$\scriptscriptstyle\smile$}}
\over D}}}^{k-1}}$ in \textbf{Step 3}.} by assuming the true channel sparsity level known at the user.
Moreover, to investigate the performance gain from the exploitation of the spatial common sparsity of CIRs, we provide the MSE performance of adaptive subspace pursuit (ASP) algorithm, which is a special case of the proposed ASSP algorithm without leveraging such spatial common sparsity of CIRs.
Simulation system parameters were set as: system carrier was $f_c=2~\rm{GHz}$, system bandwidth was $f_s=10~\rm{MHz}$, DFT size was $N= 4096$, and the length of the guard interval was $N_g=64$, which could combat the maximum delay spread of $6.4~\mu s$ \cite{3GPP,jsac_xinyu}. We consider the 4 $\times$ 16 planar antenna array ($M=64$), and $M_G= 32$ is considered to guarantee the spatial common sparsity of channels in each antenna group, the number of pilots to estimate channels for one antenna group is $N_p$, and the pilot overhead ratio is $\eta_p = (N_p M)/(N f_p M_G)$.
The International Telecommunications Union Vehicular-A (ITU-VA) channel model with $P=6$ paths was adopted \cite{3GPP}.
Finally, $p_{\rm{th}}$ was set as 0.1, 0.08, 0.06, 0.05, and 0.04 for $\rm{SNR}=10~\rm{dB}$, $15~\rm{dB}$, $20~\rm{dB}$, $25~\rm{dB}$, and $30~\rm{dB}$, respectively.
%
%For the CSI matrix over sparse multipath channels, we study its propagation conditions by examining the eigenvalue distribution of $\bf{G}\bf{G}^{\rm{H}}$. Meanwhile, the capacity of massive MIMO over sparse multipath channels is also studied, where the BCS-based CSI matrix with entries subjected to i.i.d. complex Gaussian distribution is compared. Furthermore, for the proposed channel estimation scheme, we investigate the convergence of the proposed ASSP algorithm, meanwhile, its mean square error (MSE) performance is studied, where the classical SAMP algorithm, the BSP algorithm with exact sparsity level of channels, and the exact LS method with path delays as priori information were compared. %SP algorithm and S-SAMP algorithm are provided with true sparsity of channels, and exact LS method indicates the exact path delays are known as prior information, which can be regarded as the performance bound.

%\begin{figure}[!tp]
%     \centering
%     \includegraphics[width=\columnwidth, keepaspectratio]
%     {figs/fig2_TWC2.eps}
%    \caption{MSE comparison of different channel estimation algorithms versus different numbers of pilot overhead and SNRs provided $M_G=32$.}%; (b): Proposed superimposed pilot schemewhere horizontal axis is path delays, and vertical %axis is delay gains
%     \label{fig:MSE32_1}
%      \vspace*{0.0mm}
%\end{figure}

Fig. \ref{fig:MSE64_1} compares the MSE performance of the ASSP algorithm, the oracle ASSP algorithm, the ASP algorithm, and the oracle LS algorithm over static ITU-VA channel.
In the simulation, we only consider the channel estimation for one OFDM symbol with $R=1$ and $f_p=1$.
% The MSE performance of the oracle LS method with the known path delays as the priori information is also included as the performance bound.
From Fig. \ref{fig:MSE64_1}, it can be observed that the ASP algorithm performs poorly.
The proposed ASSP algorithm outperforms the ASP algorithm, since the spatial common sparsity of MIMO channels is leveraged for the enhanced channel estimation performance.
Moreover, for $\eta_p  \ge 19.04\%$, the ASSP algorithm and the oracle ASSP algorithm have the similar MSE performance, and their performance approaches that of the oracle LS algorithm. This indicates that the proposed ASSP algorithm can reliably acquire the channel sparsity level and the support set for $\eta_p  \ge 19.04\%$.
 %since the oracle ASSP algorithm with the known sparsity level as the priori information can be considered as the benchmark to judge the sparse signal recovery performance of ASSP algorithm in practical channel estimation scenes  where the sparsity is not known.
%Second, it is clear from Fig. \ref{fig:MSE64_1} that the classical
%Furthermore, the proposed ASSP algorithm can approach  when the pilot overhead $\eta_p  = {N_p}/N>19.04\%$.
Moreover, the low pilot overhead implies that the average pilot overhead to estimate the channel associated with one transmit antenna is $N_{p\_{\rm{avg}}}=N_p/M_G=12.18$, which approaches $2P=12$, the minimum number of observations to reliably recover a $P$-sparse signal \cite{spark}. Therefore, the good sparse signal recovery performance of the proposed non-orthogonal pilot scheme and the near-optimal channel estimation performance of the proposed ASSP algorithm are confirmed.
\begin{figure}[!tp]
     \centering
    % \vspace*{-10.0mm}
     \includegraphics[width=9.0cm, keepaspectratio]
     {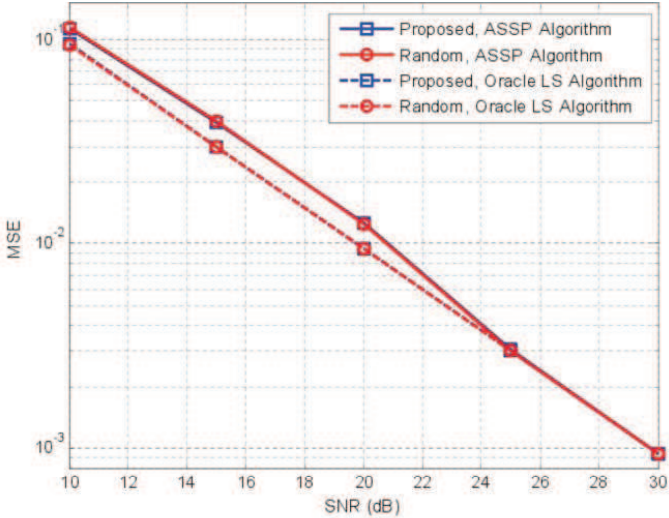}
       \vspace*{-5.0mm}
    \caption{MSE performance comparison of the proposed pilot placement scheme and the conventional random pilot
placement scheme.}%; (b): Proposed superimposed pilot schemewhere horizontal axis is path delays, and vertical %axis is delay gains
     \label{fig:random_pilot}
     \vspace*{-2.0mm}
\end{figure}

From Fig. \ref{fig:MSE64_1}, we observe that the ASSP algorithm outperforms the oracle ASSP algorithm for $\eta_p  < 19.04\%$, and its performance is even better than the performance bound obtained by the oracle LS algorithm with $N_{p\_{\rm{avg}}}<2P$ at $\rm{SNR}=10~\rm{dB}$. This is because the ASSP algorithm can adaptively acquire the effective channel sparsity level, denoted by $P_{\rm{eff}}$, instead of $P$ can be used to achieve better channel estimation performance. Consider $\eta_p =17.09 \%$ at $\rm{SNR}=10~\rm{dB}$ as an example, we can find that $P_{\rm{eff}}=5$ with high probability for the ASSP algorithm if we refer to Fig. \ref{fig:iter_64}. Hence, the average pilot overhead for each transmit antenna $N_{p\_{\rm{avg}}}=N_p/M_G=10.9$ is still larger than $2P_{\rm{eff}}=10$. From the analysis above, we come to the conclusion that, when $N_p$ is insufficient to estimate channels with $P$, the ASSP algorithm will estimate sparse channels with $P_{\rm{eff}}<P$, where path gains accounting for the majority of the channel energy will be estimated, while those with the small energy are discarded as noise. It should be pointed out that the MSE performance fluctuation of the ASSP algorithm at SNR = 10 dB is caused by the fact that $P_{\rm{eff}}$ increases from 5 to 6 when ${\eta _p}$ increases, which leads some strong noise to be estimated as the channel paths, and thus degrades the MSE performance.

%\begin{figure}[!tp]
%     \centering
%     \includegraphics[width=\columnwidth, keepaspectratio]
%     {figs/iter_num_32.eps}
%    \caption{Convergence and cognitive channel sparsity of the proposed S-SAMP algorithm with $M_G=32$.}%; (b): Proposed superimposed pilot %schemewhere horizontal axis is path delays, and vertical %axis is delay gains
%     \label{fig:iter_32}
%\end{figure}

Fig. \ref{fig:iter_64} depicts the estimated channel sparsity level of the proposed ASSP algorithm against SNR and pilot overhead ratio, where the vertical axis and the horizontal axis represent the used pilot overhead ratio and the adaptively estimated channel sparsity level, respectively, and the chroma denotes the probability of the estimated channel sparsity level. In the simulation, we consider $R=1$ and $f_p=1$ without exploiting the temporal channel correlation.
Clearly, the proposed ASSP algorithm can acquire the true channel sparsity level with high probability when SNR and pilot overhead ratio increase. Moreover, even in the case of insufficient number of pilots which cannot guarantee the reliable recovery of sparse channels, the proposed ASSP algorithm can still acquire the channel sparsity level with a slight deviation from the true channel sparsity level.

\begin{figure}[!tp]
     \centering
    % \vspace*{-10.0mm}
     \includegraphics[width=8.9cm, keepaspectratio]
     {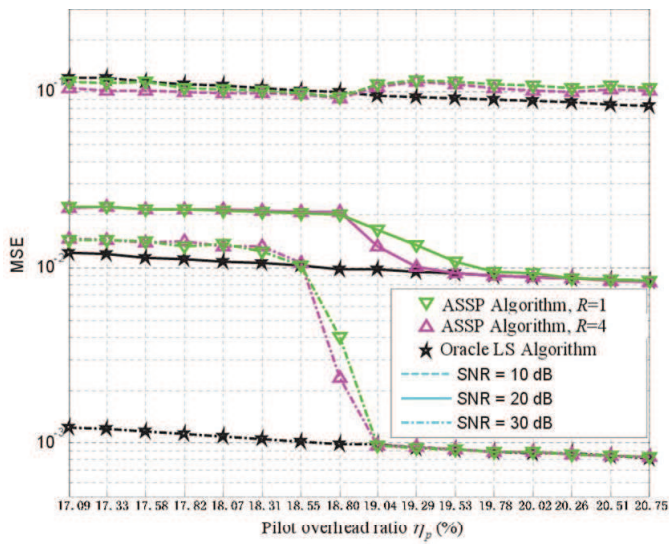}
    %   \vspace*{-6.0mm}
    \caption{MSE performance comparison of the ASSP algorithm with different $R$'s over time-varying ITU-VA channel with the mobile speed of 60 km/h.}%; (b): Proposed superimposed pilot schemewhere horizontal axis is path delays, and vertical %axis is delay gains
     \label{fig:R4_64}
   %  \vspace*{-2.0mm}
\end{figure}

Fig. \ref{fig:random_pilot} compares the MSE performance of the proposed pilot placement scheme and the conventional random pilot placement scheme \cite{CS_Sparse}, where the proposed ASSP algorithm and the oracle LS algorithm are used. In the simulation, we consider $R=1$, $f_p=1$, and ${\eta _p}=19.53~\%$. Clearly, the proposed pilot placement scheme and the conventional random pilot placement scheme have very similar performance. Due to the regular pilot placement, the proposed uniformly-spaced pilot placement scheme can be more easily implemented in practical systems. Moreover, the uniformly-spaced pilot placement scheme has been used in LTE-Advanced systems, which can facilitate massive MIMO to be backward compatible with
current cellular networks \cite{LTE_review}.

%\begin{figure}[!tp]
%     \centering
%     \includegraphics[width=\columnwidth, keepaspectratio]
%     {figs/fig4_TWC2.eps}
%    \caption{MSE comparison of S-SAMP algorithm with $M_G=32$ against different $R$'s.}%; (b): Proposed superimposed pilot schemewhere horizontal %axis is path delays, and vertical %axis is delay gains
%     \label{fig:R4_32}
%\end{figure}

Fig. \ref{fig:R4_64} provides the MSE performance comparison of the proposed ASSP algorithm with ($R=4$) and without ($R=1$) exploiting the temporal common support of wireless channels, where the time-varying ITU-VA channel with the user's mobile speed of 60 km/h is considered.
In the simulation, $f_p=1$, and $R=1$ or $4$ denotes the joint processing of the received pilot signals in $R$ successive OFDM symbols.
It can be observed that the channel estimation exploiting the temporal channel correlation performs better
than that without considering such channel property, especially at low SNR, since more
measurements can be used for the improved channel estimation performance.
%It can be observed that ASSP algorithm, which jointly estimates channels of $R=4$ OFDM symbols, works better than that with $R=1$.
%It is because the channel estimate with $R=4$ exploits the spatio-temporal common support of wireless MIMO channels, and  than that only utilizing the spatial common support of MIMO channels.
Additionally, by jointly estimating MIMO channels associated with multiple OFDM symbols, we can further reduce the required computational complexity. To be specific, the main computational burden comes from the Moore-Penrose matrix inversion operation as discussed in Section \ref{complexity}, and the joint processing of received pilot signals in $R$ OFDM symbols can share the Moore-Penrose matrix inversion operation, which indicates that the complexity can be reduced to $1/R$ of the complexity without using the temporal channel correlation.
%Meanwhile, the channel estimation by simultaneously estimating multiple channels of $R$ successive OFDM symbols can share the LS operation in each iteration due to the temporal common support of MIMO channels.
%In this way, the complexity of ASSP algorithm by exploiting temporal correlation of channels can be reduced to $1/R$ compared with that not using the temporal correlation.

\begin{figure}[!tp]
     \centering
     %\vspace*{2.0mm}
     \includegraphics[width=9cm, keepaspectratio]
     {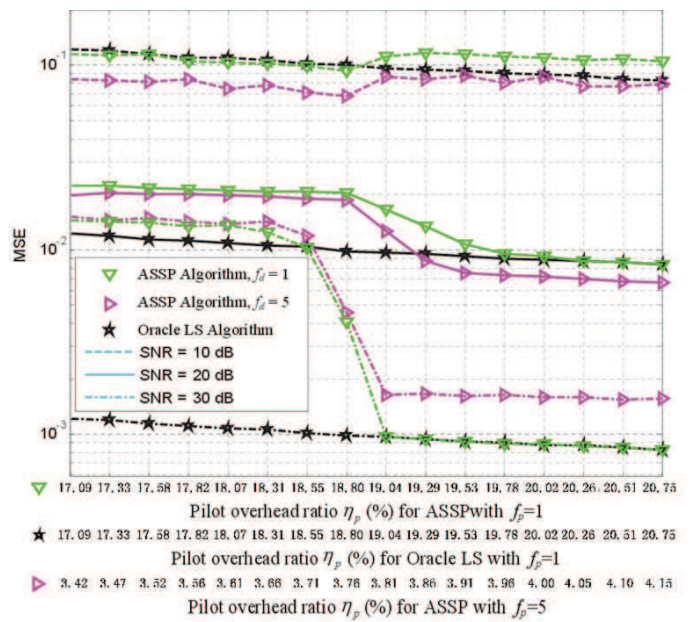}
   %   \vspace*{-8.0mm}
    \caption{MSE performance comparison of ASSP algorithm with different $f_d$'s over time-varying ITU-VA channel with the mobile speed of 60 km/h.}%; (b): Proposed superimposed pilot schemewhere horizontal axis is path delays, and vertical %axis is delay gains
     \label{fig:R5_64}
    %    \vspace*{-4.0mm}
\end{figure}

Fig. \ref{fig:R5_64} investigates the performance of the proposed space-time adaptive pilot scheme with different $f_p$'s in practical massive MIMO systems, where $R=1$, the time-varying ITU-VA channel with the user's mobile speed of $60~\rm{km/h}$ is considered, and the pilot overhead ratios with different $f_p$'s are provided.
In the simulation, $f_d=1$ and $f_d=5$ are considered, and the linear interpolation algorithm is used to estimate channels for OFDM symbols without pilots. %For the space-time adaptive pilot with $f_d>1$, the pilot overhead ratio is $\eta_p = N_p M/N f_p M_G$.
%Besides, the proposed channel estimation sheme without using the space-time division based superimposed pilot strategy is also plotted as the benchmark.
From Fig. \ref{fig:R5_64}, it can be observed that the case with $f_d=5$ only suffers from a negligible performance loss compared to that with $f_d=1$ at $\rm{SNR}=30~\rm{dB}$. While for $\rm{SNR}\le 20~\rm{dB}$,  the case with $f_d=5$ is better than that with $f_d=1$, since the linear interpolation can reduce the effective noise. %It should be noted that by using the proposed space-time division based superimposed pilot strategy, the total pilot overhead can be reduced significantly, where we consider $M=128$ as example.
By exploiting the temporal channel correlation, the proposed space-time adaptive pilot scheme can substantially reduce the required pilot overhead for channel estimation without the obvious performance loss.%, and thus the spectrum and energy resources for effective data transmission can be significantly increased.

\begin{figure}[!tp]
     \centering
     %\vspace*{-2.0mm}
     \includegraphics[width=9cm, keepaspectratio]
     {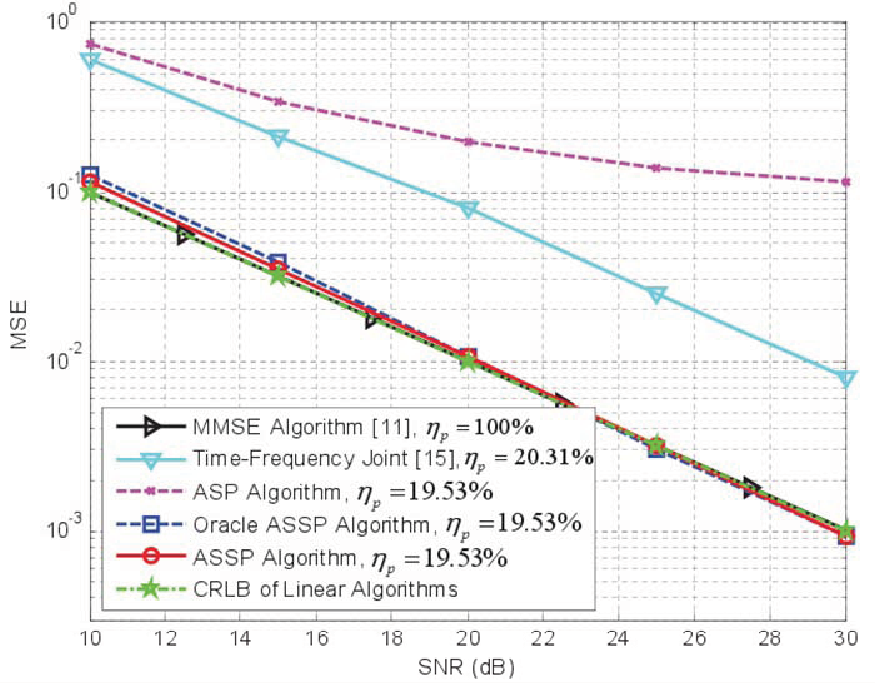}
    % \vspace*{-2.3mm}
    \caption{MSE performance comparison of different channel estimation schemes for FDD massive MIMO systems. }%; (b): Proposed superimposed pilot schemewhere horizontal axis is path delays, and vertical %axis is delay gains
     \label{fig:final}
     % \vspace*{-1.7mm}
\end{figure}

Fig. \ref{fig:final} provides the MSE performance comparison of several channel estimation schemes for FDD massive MIMO systems, where we consider the channel estimation for one OFDM symbol with $R=1$ and $f_p=1$. The Cramer-Rao lower bound (CRLB) of conventional linear channel estimation schemes (e.g., minimum mean square error (MMSE) algorithm and LS algorithm) is also plotted as the performance benchmark, where ${\mathop{\rm CRLB}} = 1/{\rm{SNR}}$~\cite{CS_dai}. The ASP algorithm does not perform well due to the insufficient pilots. The time-frequency joint training based scheme \cite{CS_dai} works poorly since the mutual interferences of time-domain training sequences of different transmit antennas degrade the channel estimation performance when $M$ is large. Both the MMSE algorithm \cite{TSP_optimal} and the proposed ASSP algorithm achieves 9 dB gain over the scheme proposed in~\cite{CS_dai}, and both of them approach the CRLB of conventional linear algorithms. It is worth mentioning that the proposed scheme enjoys the significantly reduced pilot overhead compared with the MMSE algorithm, since the MMSE algorithm work well only when (\ref{equ:compact2}) is well-determined or over-determined. Finally, since the proposed ASSP algorithm can adaptively acquire the channel sparsity level and discards the multipath components buried by the noise at low SNR for improved channel estimation, we can find the proposed scheme even works better than the oracle ASSP algorithm at low SNR.

\begin{figure}[!tp]
     \centering
     %\vspace*{-10.0mm}
     \includegraphics[width=9cm, keepaspectratio]%[angle=-90]
     {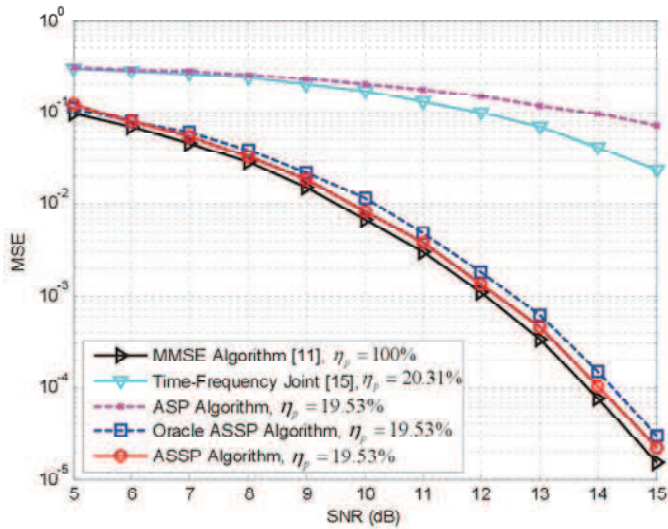}
  %   \vspace*{-5.0mm}
    \caption{BER performance comparison of different channel estimation schemes for FDD massive MIMO systems. }%; (b): Proposed superimposed pilot schemewhere horizontal axis is path delays, and vertical %axis is delay gains
     \label{fig:BER}
   %   \vspace*{-2.0mm}
\end{figure}

\begin{figure}[!tp]
     \centering
     %\vspace*{-10.0mm}
     \includegraphics[width=9cm, keepaspectratio]
     {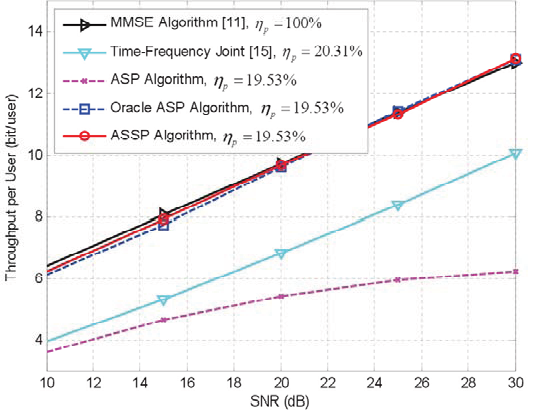}
    % \vspace*{-5.0mm}
    \caption{Comparison of average achievable throughput per user of different channel estimation schemes for FDD massive MIMO systems. }%; (b): Proposed superimposed pilot schemewhere horizontal axis is path delays, and vertical %axis is delay gains
     \label{fig:capacity}
     % \vspace*{-2.0mm}
\end{figure}

Fig. \ref{fig:BER} and Fig. \ref{fig:capacity} compare the downlink bit error rate (BER) performance and average achievable throughput per user, respectively, where the BS using zero-forcing (ZF) precoding is assumed to know the estimated downlink channels. In the simulations, the BS with $M=64$ antennas simultaneously serves $K=8$ users using 16-QAM, and the ZF precoding is based on the estimated channels corresponding to Fig. \ref{fig:final} under the same setup. It can be observed that the proposed channel estimation scheme outperforms its counterparts.

%some path gains of small energy can be negligible compared with AWGN at low SNR, thus the effective sparsity of wireless channels could be different from the true sparsity. Therefore, sparse channel estimation with effective sparsity $P_{\rm{eff}}$ could acquire better performance than that of true sparsity $P$.

%it has been shown that for an high-dimension sparse signal,  \cite{SPM}

\begin{figure}[!tp]
     \centering
     %\vspace*{-10.0mm}
     \includegraphics[width=8.9cm, keepaspectratio]
     {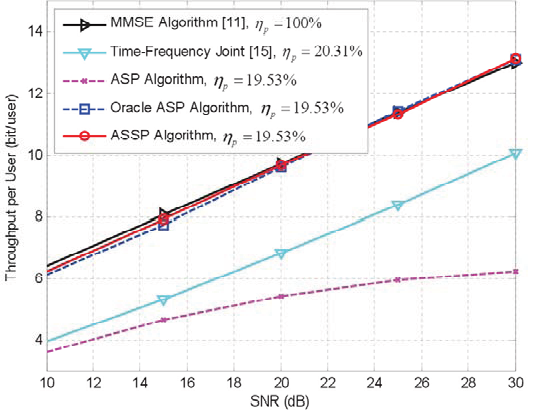}
     %\vspace*{-5.0mm}
    \caption{Comparison of average achievable throughput per user of different pilot decontamination schemes for multi-cell FDD massive MIMO systems. }%; (b): Proposed superimposed pilot schemewhere horizontal axis is path delays, and vertical %axis is delay gains
     \label{fig:capacity_multi}
      %\vspace*{-2.0mm}
\end{figure}

Fig. \ref{fig:capacity_multi} compares the average achievable throughput per user of different pilot decontamination schemes. In the simulations, we consider a multi-cell massive MIMO system with ${\cal{L}}=7$, $M=64$, $K=8$ sharing the same bandwidth, where the average achievable throughput per user in the central target cell suffering from the pilot contamination is investigated. Moreover, we consider $R=1$, $f_d=7$, the path loss factor is 3.8 dB/km, the cell radius is 1 km, the distance ${\cal{D}}$ between the BS and its users can be from 100 m to 1 km, the SNR (the power of the unprecoded signal from the BS is considered in SNR) for cell-edge user is 10 dB, the mobile speed of users is 3 km/h. The BSs using zero-forcing (ZF) precoding is assumed to know the estimated downlink channels achieved by the proposed ASSP algorithm. For the FDM scheme, pilots of ${\cal{L}}=7$ cells are orthogonal in the frequency domain. The optimal performance is achieved by the FDM scheme when the users are static. Pilots of ${\cal{L}}=7$ cells in TDM are transmitted in ${\cal{L}}=7$ successive different time slots. In TDM scheme, the channel estimation of users in central target cells suffers from the precoded downlink data transmission of other cells, where two cases are considered. The ``cell-edge" case indicates that when users in the central target cell estimate the channels, the precoded downlink data transmission in other cells can guarantee SNR = 10 dB for their cell-edge users. While the ``ergodic" case indicates that when users in the central target cell estimate the channels, the precoded downlink data transmission in other cells can guarantee SNR = 10 dB for their users with the the ergodic distance ${\cal{D}}$ from 100 m to 1 km. The negligible performance gap between the FDM scheme and the optimal one is due to the variation of time-varying channels, but it suffers from the high pilot overhead. The TDM scheme with the ``cell-edge" case performs worst. While the performance of the TDM scheme with the ``ergodic" case approaches that of the optimal one. The simulation results in Fig. \ref{fig:capacity_multi} indicates that the TDM scheme with low pilot overhead can achieve the good performance when dealing the pilot contamination in multi-cell FDD massive MIMO systems. Moreover, if some appropriate scheduling strategies are considered \cite{multicell}, the performance of the TDM scheme can be further improved.

%\vspace*{-2.0mm}
\section{Conclusions}
In this paper, we have proposed the SCS-based spatio-temporal joint channel estimation scheme for FDD massive MIMO systems, whereby the intrinsically spatio-temporal common sparsity of wireless MIMO channels is exploited to reduce the pilot overhead. First, the non-orthogonal pilot scheme at the BS and the ASSP algorithm at the user can reliably estimate channels with significantly reduced pilot overhead. Then, the space-time adaptive pilot scheme can further reduce the required pilot overhead according to the mobility of users. Moreover, we discuss the proposed
channel estimation scheme in multi-cell scenario. Additionally, we discuss the non-orthogonal pilot design to achieve the reliable channel estimation under the framework of CS theory, and the convergence analysis as well as the complexity analysis of the proposed ASSP algorithm are also provided. Simulation results have shown that the proposed channel estimation scheme can achieve much better channel estimation performance than its counterparts with substantially reduced pilot overhead, and it only suffers from a negligible performance loss when compared with the performance bound.
\vspace*{-2.0mm}

\appendix
%\vspace*{-2.0mm}
\subsection{Proof of Theorem~\ref{T1}}\label{TPa}

We first provide the definition of SRIP for ${\bf{\Psi }}$ in our problem ${\bf{Y}} = {\bf{\Psi D}}+{\bf{W}}$ (\ref{equ:common}), where $\bf{D}$ has the structured sparsity as illustrated in (\ref{equ:common3}). Particularly, the SRIP can be expressed as
 \begin{equation}\label{equ:app003}
 \begin{small}
\sqrt {1 - \delta } \left\| {\bf{D}}_\Omega \right\|_F^{} \le \left\| {{{\bf{\Psi }}_\Omega }{\bf{D}}_\Omega} \right\|_F \le \sqrt {1 + \delta } \left\| {\bf{D}}_\Omega \right\|_F^{},
 \end{small}
\end{equation}
where $\delta  \in \left[ {0,1} \right)$, ${\Omega}$ is an arbitrary set with ${\left| {\Omega} \right|_c} \le P$, and ${\delta _P}$ is the infimum of all $\delta$ satisfying (\ref{equ:app003}). Note that for (\ref{equ:app003}), ${\bf{\Psi }} = \left[ {{{\bf{\Psi }}_1},{{\bf{\Psi }}_2},\cdots ,{{\bf{\Psi }}_{L}}} \right]\in \mathbb{C}^{{N_p} \times ML}$ with ${{\bf{\Psi }}_l} \in \mathbb{C}^{{N_p} \times M} $ for $1\le l \le L$, ${\bf{D}} = {\rm{ }}{[{\bf{D}}_1^{\rm{T}},{\bf{D}}_2^{\rm{T}}, \cdots ,{\bf{D}}_L^{\rm{T}}]^{\rm{T}}}\in \mathbb{C}^{ML \times {R}}$ with ${\bf{D}}_l\in \mathbb{C}^{M \times {R}}$ for $1\le l \le L$, ${{\bf{\Psi }}_{ \Omega}} = {\left[ {{\bf{\Psi }}_{{ \Omega (1)}}^{\rm{}},{\bf{\Psi }}_{{ \Omega (2)}}^{\rm{}}, \cdots ,{\bf{\Psi }}_{{ \Omega ({|  \Omega  |_c})}}^{\rm{}}} \right]^{\rm{}}}$ and ${{\bf{ D}}_{ \Omega}} = {\left[ {{\bf{D}}_{{ \Omega (1)}}^{\rm{T}},{\bf{D}}_{{ \Omega (2)}}^{\rm{T}}, \cdots ,{\bf{D}}_{{ \Omega ({|  \Omega  |_c})}}^{\rm{T}}} \right]^{\rm{T}}}$, and ${ \Omega(1)} < { \Omega (2)} <  \cdots  < { \Omega ({|  \Omega  |_c})}$ are elements in the set $ \Omega$.
Clearly, for two different sparsity levels $P_1$ and $P_2$ with $P_1 < P_2$, we have $\delta_{P_1} \le \delta_{P_2}$. Moreover, for two sets with ${\Omega _1} \cap {\Omega _2} = \phi $ and the structured sparse matrix $\bf{D}$ with the support set ${\Omega_2}$, we have
\begin{equation}\label{equ:app0031}
\begin{small}
{\left\| {{\bf{\Psi }}_{{\Omega _1}}^{\rm{H}}{{\bf{\Psi }}}{\bf{D}}} \right\|_F} ={\left\| {{\bf{\Psi }}_{{\Omega _1}}^{\rm{H}}{{\bf{\Psi }}_{{\Omega _2}}}{\bf{D}}_{{\Omega _2}}} \right\|_F} \le {\delta _{{{\left| {{\Omega _1}} \right|}_c} + {{\left| {{\Omega _2}} \right|}_c}}}{\left\| {\bf{D}} \right\|_F},
\end{small}
\end{equation}
\begin{equation}\label{equ:app004}
\begin{small}
\begin{array}{l}
\!\!\!\!(1 - \frac{{{\delta _{{{\left| {{\Omega _1}} \right|}_c} + {{\left| {{\Omega _2}} \right|}_c}}}}}{{\sqrt {(1 - {\delta _{{{\left| {{\Omega _1}} \right|}_c}}})(1 - {\delta _{{{\left| {{\Omega _2}} \right|}_c}}})} }}){\left\| {{{\bf{\Psi }}_{{\Omega _2}}}{{\bf{D}}_{{\Omega _2}}}} \right\|_F}\\~~~~~~~~
\le {\left\| {({\bf{I}}{\rm{ - }}{{\bf{\Psi }}_{{\Omega _1}}}{\bf{\Psi }}_{{\Omega _1}}^\dag ){{\bf{\Psi }}_{{\Omega _2}}}{{\bf{D}}_{{\Omega _2}}}} \right\|_F} \le {\left\| {{{\bf{\Psi }}_{{\Omega _2}}}{{\bf{D}}_{{\Omega _2}}}} \right\|_F},
\end{array}
\end{small}
\end{equation}
which will be proven in Appendix~\ref{TPb} and~\ref{TPc}, respectively.

To prove (\ref{equ:app001}), we need to investigate the upper bound of ${\left\| {{\bf{D}}{\rm{ - }}{\bf{\hat D}}} \right\|_F}$, which can be expressed as
{\setlength\abovedisplayskip{1pt plus 3pt minus 7pt}
\setlength\belowdisplayskip{1pt plus 3pt minus 7pt}
\begin{equation}\label{equ:16}
\begin{small}
\begin{array}{l}
\!\!\!\!\!\!\!\!\!\!{\left\| {{\bf{D}} - {\bf{\hat D}}} \right\|_F} \le {\left\| {{{\bf{D}}_{\hat \Omega }} - {\bf{\Psi }}_{\hat \Omega }^\dag {\bf{Y}}} \right\|_F} + {\left\| {{{\bf{D}}_{{\Omega _T}/\hat \Omega }}} \right\|_F}\\
~~~~~~~~~= {\left\| {{{\bf{D}}_{\hat \Omega }} - {\bf{\Psi }}_{\hat \Omega }^\dag ({{\bf{\Psi }}_{{\Omega _T}}}{{\bf{D}}_{{\Omega _T}}} + {\bf{W}})} \right\|_F} + {\left\| {{{\bf{D}}_{{\Omega _T}/\hat \Omega }}} \right\|_F}\\
~~~~~~~~~\le {\left\| {{{\bf{D}}_{\hat \Omega }} - {\bf{\Psi }}_{\hat \Omega }^\dag {{\bf{\Psi }}_{{\Omega _T}}}{{\bf{D}}_{{\Omega _T}}}} \right\|_F} + {\left\| {{\bf{\Psi }}_{\hat \Omega }^\dag {\bf{W}}} \right\|_F} + {\left\| {{{\bf{D}}_{{\Omega _T}/\hat \Omega }}} \right\|_F}\\
~~~~~~~~~= {\left\| {{\bf{\Psi }}_{\hat \Omega }^\dag {{\bf{\Psi }}_{{\Omega _T}/\hat \Omega }}{{\bf{D}}_{{\Omega _T}/\hat \Omega }}} \right\|_F} + {\left\| {{\bf{\Psi }}_{\hat \Omega }^\dag {\bf{W}}} \right\|_F} + {\left\| {{{\bf{D}}_{{\Omega _T}/\hat \Omega }}} \right\|_F},%\\
%\le {\left\| {{\bf{\Psi }}_{\hat \Omega }^\dag {{\bf{\Psi }}_{{\Omega _T}/\hat \Omega }}} \right\|_F}{\left\| {{{\bf{D}}_{{\Omega _T}/\hat \Omega }}} \right\|_F} + {\left\| {{\bf{\Psi }}_{\hat \Omega }^\dag } \right\|_F}{\left\| {\bf{W}} \right\|_F} + {\left\| {{{\bf{D}}_{{\Omega _T}/\hat \Omega }}} \right\|_F},
\end{array}
\end{small}
\end{equation}
}where $\hat \Omega $ is the estimated support set, $\Omega_{T} $ is the correct support set, and ${{\Omega _T}/\hat \Omega }$ denotes a set whose elements belong to ${\Omega _T}$ except for ${\hat  \Omega }$. The first inequality is due to $\left\| {\bf{D}} \right\|_F^2 = \left\| {{{\bf{D}}_{\hat \Omega }}} \right\|_F^2 + \left\| {{{\bf{D}}_{{\Omega _T}/\hat \Omega }}} \right\|_F^2$. The second equality is due to ${{\bf{\Psi }}_{{\Omega _T}}}{{\bf{D}}_{{\Omega _T}}} = {{\bf{\Psi }}_{{\Omega _T}/\hat \Omega }}{{\bf{D}}_{{\Omega _T}/\hat \Omega }} + {{\bf{\Psi }}_{{\Omega _T} \cap \hat \Omega }}{{\bf{D}}_{{\Omega _T} \cap \hat \Omega }}$ and ${{\bf{D}}_{\hat \Omega }} = {\bf{\Psi }}_{\hat \Omega }^\dag {{\bf{\Psi }}_{{\Omega _T} \cap \hat \Omega }}{{\bf{D}}_{{\Omega _T} \cap \hat \Omega }}$.

For ${\left\| {{\bf{\Psi }}_{\hat \Omega }^\dag {{\bf{\Psi }}_{{\Omega _T}/\hat \Omega }}{{\bf{D}}_{{\Omega _T}/\hat \Omega }}} \right\|_F} $, we have
\begin{equation}\label{equ:16.10}
\begin{small}
\begin{array}{l}
\!\!\!\!\!\!\!\!\!{\left\| {{\bf{\Psi }}_{\hat \Omega }^\dag {{\bf{\Psi }}_{{\Omega _T}/\hat \Omega }}}{{\bf{D}}_{{\Omega _T}/\hat \Omega }} \right\|_F} \\
= {\left\| {{{({\bf{\Psi }}_{\hat \Omega }^{\rm{H}}{\bf{\Psi }}_{\hat \Omega }^{})}^{ - 1}}{\bf{\Psi }}_{\hat \Omega }^{\rm{H}}{{\bf{\Psi }}_{{\Omega _T}/\hat \Omega }}}{{\bf{D}}_{{\Omega _T}/\hat \Omega }} \right\|_F}\le \frac{{{\delta _{2P}}}}{{1 - {\delta _P}}} {\left\| {{\bf{D}}_{{\Omega _T}/\hat \Omega }} \right\|_F},
\end{array}
\end{small}
\end{equation}
where the inequality of (\ref{equ:16.10}) is due to (\ref{equ:app003}) and (\ref{equ:app0031}). Similarly, we have ${\left\| {{\bf{\Psi }}_{\hat \Omega }^\dag }{\bf{W}} \right\|_F} \le \frac{{\sqrt {1 + {\delta _P}} }}{{1 - {\delta _P}}}{\left\|{\bf{W}} \right\|_F}$. Thus we have
\begin{equation}\label{equ:16.1}
\begin{small}
\begin{array}{l}
\!\!\!\!\!\!\!\!{\left\| {{\bf{D}} - {\bf{\hat D}}} \right\|_F}\le \frac{{1 - {\delta _P} + {\delta _{2P}}}}{{1 - {\delta _P}}}{\left\| {{{\bf{D}}_{{\Omega _T}/\hat \Omega }}} \right\|_F} + \frac{{\sqrt {1 + {\delta _P}} }}{{1 - {\delta _P}}}{\left\| {\bf{W}} \right\|_F}.
\end{array}
\end{small}
\end{equation}
Then we will investigate the relationship between ${\left\| {{{\bf{D}}_{{\Omega _T}/\hat \Omega }}} \right\|_F}$ and ${\left\| {\bf{W}} \right\|_F}$. It should be pointed out that, after we get $\hat \Omega$, we have ${\left\| {{{\bf{R}}^{k - 1}}} \right\|_F} \le {\left\| {\bf{R}}^{k} \right\|_F}$, % and $\hat \Omega = {{\tilde \Omega }^{k - 1}}={{\tilde \Omega }^{k }}$,
which inspires us to first study the relationship between ${\left\| {{{\bf{R}}^k}} \right\|_F}$ and ${\left\| {{{\bf{R}}^{k-1}}} \right\|_F}$.

For ${\left\| {{{\bf{R}}^k}} \right\|_F}$, we can obtain
 \begin{equation}\label{equ:app0.1}
 \begin{small}
\begin{array}{l}
{\left\| {{{\bf{R}}^k}} \right\|_F} = {\left\| {{\bf{\Psi D}}{\rm{ + }}{\bf{W}}{\rm{ - }}{{\bf{\Psi }}_{{{\tilde \Omega }^k}}}{\bf{\Psi }}_{{{\tilde \Omega }^k}}^\dag ({\bf{\Psi D}}{\rm{ + }}{\bf{W}})} \right\|_F}\\
~~~~ \le {\left\| {({\bf{I}}{\rm{ - }}{{\bf{\Psi }}_{{{\tilde \Omega }^k}}}{\bf{\Psi }}_{{{\tilde \Omega }^k}}^\dag ){{\bf{\Psi }}_{{\Omega _T}/{{\tilde \Omega }^k}}}{{\bf{D}}_{{\Omega _T}/{{\tilde \Omega }^k}}}} \right\|_F}{\rm{ + }}{\left\| {{\bf{W}}{\rm{ - }}{{\bf{\Psi }}_{{{\tilde \Omega }^k}}}{\bf{\Psi }}_{{{\tilde \Omega }^k}}^\dag {\bf{W}}} \right\|_F}\\
~~~~ \le {\left\| {{{\bf{\Psi }}_{{\Omega _T}/{{\tilde \Omega }^k}}}{{\bf{D}}_{{\Omega _T}/{{\tilde \Omega }^k}}}} \right\|_F}{\rm{ + }}{\left\| {\bf{W}} \right\|_F}\\~~~~
\le \sqrt {1{\rm{ + }}{\delta _P}} {\left\| {{{\bf{D}}_{{\Omega _T}/{{\tilde \Omega }^k}}}} \right\|_F}{\rm{ + }}{\left\| {\bf{W}} \right\|_F}.
\end{array}
\end{small}
\end{equation}
where we have ${\bf{\Psi D}}{\rm{ = }}{{\bf{\Psi }}_{{\Omega _T} \cap {{\tilde \Omega }^k}}}{{\bf{D}}_{{\Omega _T} \cap {{\tilde \Omega }^k}}}{\rm{ + }}{{\bf{\Psi }}_{{\Omega _T}/{{\tilde \Omega }^k}}}{{\bf{D}}_{{\Omega _T}/{{\tilde \Omega }^k}}}$, ${{{\bf{\Psi }}_{{\Omega _T} \cap {{\tilde \Omega }^k}}}{{\bf{D}}_{{\Omega _T} \cap {{\tilde \Omega }^k}}}{\rm{ = }}{{\bf{\Psi }}_{{{\tilde \Omega }^k}}}{\bf{\Psi }}_{{{\tilde \Omega }^k}}^\dag {{\bf{\Psi }}_{{\Omega _T} \cap {{\tilde \Omega }^k}}}{{\bf{D}}_{{\Omega _T} \cap {{\tilde \Omega }^k}}}}$, and the second inequality is due to (\ref{equ:app004}) and ${\left\| {{\bf{W}}{\rm{ - }}{{\bf{\Psi }}_{{{\tilde \Omega }^k}}}{\bf{\Psi }}_{{{\tilde \Omega }^k}}^\dag {\bf{W}}} \right\|_F}\le {\left\| {\bf{W}} \right\|_F}$.

On the other hand, we consider ${\left\| {{{\bf{R}}^{k-1}}} \right\|_F}$, which can be expressed as
 \begin{equation}\label{equ:app0.2}
 \begin{small}
\begin{array}{l}
\!\!\!{\left\| {{{\bf{R}}^{k{\rm{ - }}1}}} \right\|_F} \ge {\left\|({\bf{I}}{\rm{ - }}{\bf{\Psi}}_{{{\tilde \Omega }^k}}{{\bf \Psi }}_{{{\tilde \Omega }^k}}^\dag ) {{{\bf{\Psi }}_{{\Omega _T}/{{\tilde \Omega }^{k{\rm{ - }}1}}}}{{\bf{D}}_{{\Omega _T}/{{\tilde \Omega }^{k{\rm{ - }}1}}}}} \right\|_F}{\rm{ - }}{\left\| {\bf{W}} \right\|_F}\\
~~~~~~~~~~~ \ge \frac{{1{\rm{ - }}{\delta _{P}}{\rm{ - }}{\delta _{2P}}}}{{1{\rm{ - }}{\delta _{P}}}}{\left\| {{{\bf{\Psi }}_{{\Omega _T}/{{\tilde \Omega }^{k{\rm{ - }}1}}}}{{\bf{D}}_{{\Omega _T}/{{\tilde \Omega }^{k{\rm{ - }}1}}}}} \right\|_F}{\rm{ - }}{\left\| {\bf{W}} \right\|_F}\\~~~~~~~~~~~
 \ge \frac{{1{\rm{ - }}{\delta _{P}}{\rm{ - }}{\delta _{2P}}}}{{\sqrt {1{\rm{ - }}{\delta _{P}}} }}{\left\| {{{\bf{D}}_{{\Omega _T}/{{\tilde \Omega }^{k{\rm{ - }}1}}}}} \right\|_F}{\rm{ - }}{\left\| {\bf{W}} \right\|_F},
\end{array}
\end{small}
\end{equation}
where the second inequality is due to (\ref{equ:app004}).

To further investigate the relationship between (\ref{equ:app0.1}) and (\ref{equ:app0.2}), we will derive the relationship between ${\left\| {{{\bf{D}}_{{\Omega _T}/{{\tilde \Omega }^{k{\rm{ }}}}}}} \right\|_F}$ and ${\left\| {{{\bf{D}}_{{\Omega _T}/{{\tilde \Omega }^{k{\rm{ - }}1}}}}} \right\|_F}$. For convenience, we denote $\Omega_\Delta   = {\Pi ^s}\left( {\left\{ {{{\left\| {{{\bf{Z}}_l}} \right\|}_F}} \right\}_{l = 1}^L} \right)$ in \textbf{Step 2.3} of Algorithm 1, then we can get
 \begin{equation}\label{equ:app1.1}
 \begin{small}
\begin{array}{l}
\!\!\!\!{\left\| {\left. {{\bf{\Psi }}_{{\Omega _\Delta }}^{\rm{H}}{\bf{R}}_{}^{k - 1}} \right\|} \right._F} ={\left\| {{\bf{\Psi }}_{{\Omega _\Delta}}^{\rm{H}}({\bf{Y}} - {\bf{\Psi }}_{{{\tilde \Omega }^{k - 1}}}^{}{\bf{\Psi }}_{{{\tilde \Omega }^{k - 1}}}^\dag {\bf{Y}})} \right\|_F}\\~~~~~~
= {\left\| {{\bf{\Psi }}_{{\Omega _\Delta}}^{\rm{H}}({\bf{\Psi D}} + {\bf{W}} - {\bf{\Psi }}_{{{\tilde \Omega }^{k - 1}}}^{}{\bf{\Psi }}_{{{\tilde \Omega }^{k - 1}}}^\dag ({\bf{\Psi D}} + {\bf{W}}))} \right\|_F}\\
~~~~~~ \le {\left. {\left\| {{\bf{\Psi }}_{{\Omega _\Delta}}^{\rm{H}}({\bf{\Psi D}} - {\bf{\Psi }}_{{{\tilde \Omega }^{k - 1}}}^{}{\bf{\Psi }}_{{{\tilde \Omega }^{k - 1}}}^\dag {\bf{\Psi D}})} \right.} \right\|_F}\\ ~~~~~~~~~~~ ~~~~~~~~~~~
 + {\left. {\left\| {{\bf{\Psi }}_{{\Omega _\Delta}}^{\rm{H}}({\bf{W}} - {\bf{\Psi }}_{{{\tilde \Omega }^{k - 1}}}^{}{\bf{\Psi }}_{{{\tilde \Omega }^{k - 1}}}^\dag {\bf{W}})} \right.} \right\|_F}.
\end{array}
 \end{small}
\end{equation}
%where the first equality is due to the definition of the residue.
For the first part of the right-hand in the inequality of (\ref{equ:app1.1}), we denote ${\bf{R'}}_{}^{k - 1} = {\bf{\Psi D}} - {\bf{\Psi }}_{{{\tilde \Omega }^{k - 1}}}^{}{\bf{\Psi }}_{{{\tilde \Omega }^{k - 1}}}^\dag {\bf{\Psi D}}$, and
\begin{equation}
\begin{small}\label{equ:app2}
\begin{array}{l}
\!\!\!{\bf{R'}}_{}^{k - 1}
{\rm{ = (}}{\bf{I}}{\rm{ - }}{\bf{\Psi }}_{{{\tilde \Omega }^{k - 1}}}^{}{\bf{\Psi }}_{{{\tilde \Omega }^{k - 1}}}^\dag )({{\bf{\Psi }}_{{\Omega _T}/{{\tilde \Omega }^{k - 1}}}}{{\bf{D}}_{{\Omega _T}/{{\tilde \Omega }^{k - 1}}}}\\~~~~~~~~~~~~~~~~~~~~~~~~~~~
 + {{\bf{\Psi }}_{{\Omega _T} \cap {{\tilde \Omega }^{k - 1}}}}{{\bf{D}}_{{\Omega _T} \cap {{\tilde \Omega }^{k - 1}}}})\\
~~~~~~ = [{{\bf{\Psi }}_{{\Omega _T}/{{\tilde \Omega }^{k - 1}}}},{\bf{\Psi }}_{{{\tilde \Omega }^{k - 1}}}^{}]
 \left[ {\begin{array}{*{20}{c}}
{{\bf{D}}_{{\Omega _T}/{{\tilde \Omega }^{k - 1}}}^{}}\\
{ - {\bf{\Psi }}_{{{\tilde \Omega }^{k - 1}}}^\dag {{\bf{\Psi }}_{{\Omega _T}/{{\tilde \Omega }^{k - 1}}}}{{\bf{D}}_{{\Omega _T}/{{\tilde \Omega }^{k - 1}}}}}
\end{array}} \right]\\
~~~~~~= {{\bf{\Psi }}_{{\Omega _T} \cup {{\tilde \Omega }^{k - 1}}}}{\bf{\tilde D}}_{}^{k - 1},
\end{array}
\end{small}
\end{equation}
where ${\bf{\Psi }}_{{\Omega _T} \cup {{\tilde \Omega }^{k - 1}}}^{} = [{\bf{\Psi }}_{{\Omega _T}/{{\tilde \Omega }^{k - 1}}}^{},{\bf{\Psi }}_{{{\tilde \Omega }^{k - 1}}}^{}]$ and ${\bf{\tilde D}}_{}^{k - 1} = {[{\bf{D}}_{{\Omega _T}/{{\tilde \Omega }^{k - 1}}}^{\rm{T}}, - {({\bf{\Psi }}_{{{\tilde \Omega }^{k - 1}}}^\dag {{\bf{\Psi }}_{{\Omega _T}/{{\tilde \Omega }^{k - 1}}}}{{\bf{D}}_{{\Omega _T}/{{\tilde \Omega }^{k - 1}}}})^{\rm{T}}}]^{\rm{T}}}$. The second equality of (\ref{equ:app2}) is due to ${{\bf{\Psi }}_{{\Omega _T} \cap {{\tilde \Omega }^{k - 1}}}}{{\bf{D}}_{{\Omega _T} \cap {{\tilde \Omega }^{k - 1}}}} - {\bf{\Psi }}_{{{\tilde \Omega }^{k - 1}}}^{}{\bf{\Psi }}_{{{\tilde \Omega }^{k - 1}}}^\dag {{\bf{\Psi }}_{{\Omega _T} \cap {{\tilde \Omega }^{k - 1}}}}{{\bf{D}}_{{\Omega _T} \cap {{\tilde \Omega }^{k - 1}}}} = {\bf{0}}$. It should be pointed out that if ${\bf{W}}={\bf{0}}$, we have ${\bf{R'}}_{}^{k - 1} = {\bf{R}}_{}^{k - 1}$. For the second part of the right-hand in the inequality of (\ref{equ:app1.1}), we have
%{\setlength\abovedisplayskip{6pt}
%\setlength\belowdisplayskip{6pt}
\begin{equation}
\begin{small}\label{equ:app2.1}
\begin{array}{l}
\!\!\!\!\!\!\!\!\!\!{\left. {\left\| {{\bf{\Psi }}_{{\Omega _\Delta }}^{\rm{H}}({\bf{W}} - {\bf{\Psi }}_{{{\tilde \Omega }^{k - 1}}}^{}{\bf{\Psi }}_{{{\tilde \Omega }^{k - 1}}}^\dag {\bf{W}})} \right.} \right\|_F}\\~
= {\left. {\left\| {{\bf{\Psi }}_{{\Omega _\Delta }}^{\rm{H}}({\bf{I}} - {\bf{\Psi }}_{{{\tilde \Omega }^{k - 1}}}^{}{\bf{\Psi }}_{{{\tilde \Omega }^{k - 1}}}^\dag ){\bf{W}}} \right.} \right\|_F} \le \sqrt {1 + {\delta _P}} {\left\| {\bf{W}} \right\|_F}.
\end{array}
\end{small}
\end{equation}
%}
By substituting (\ref{equ:app2}) and (\ref{equ:app2.1}) into (\ref{equ:app1.1}), we have
%{\setlength\abovedisplayskip{6pt}
%\setlength\belowdisplayskip{6pt}
\begin{equation}\label{equ:app3}
\begin{small}
\begin{array}{l}
{\left\| {\left. {{\bf{\Psi }}_{{\Omega _\Delta }}^{\rm{H}}{\bf{R}}_{}^{k - 1}} \right\|} \right._F} %= {\left\| {{\bf{\Psi }}_{{\Omega _\Delta }}^{\rm{H}}({\bf{Y}} - {\bf{\Psi }}_{{{\tilde \Omega }^{k - 1}}}^{}{\bf{\Psi }}_{{{\tilde \Omega }^{k - 1}}}^\dag {\bf{Y}})} \right\|_F}\\
 \le {\left\| {{\bf{\Psi }}_{{\Omega _\Delta }}^{\rm{H}}{{\bf{\Psi }}_{{\Omega _T} \cup {{\tilde \Omega }^{k - 1}}}}{\bf{\tilde D}}_{}^{k - 1}} \right\|_F} + \sqrt {1 + {\delta _P}} {\left\| {\bf{W}} \right\|_F}\\
~~~~~~~~~~~~~~~~~~~ = {\left\| {{\bf{\Psi }}_{{\Omega _\Delta }}^{\rm{H}}} {\bf{R'}}_{}^{k - 1} \right\|_F} + \sqrt {1 + {\delta _P}} {\left\| {\bf{W}} \right\|_F},
\end{array}
\end{small}
\end{equation}
%}
%where the inequality is due to (\ref{equ:app0031}).
On the other hand, we have
%{\setlength\abovedisplayskip{6pt}
%\setlength\belowdisplayskip{6pt}
 \begin{equation}\label{equ:app3.1}
 \begin{small}
\begin{array}{l}
\!\!\!\!\!\!\!\!\!\!\!\!\!{\left\| {\left. {{\bf{\Psi }}_{{\Omega _\Delta }}^{\rm{H}}{\bf{R}}_{}^{k - 1}} \right\|} \right._F} \mathop  \ge \limits^{ }{\left\| {{\bf{\Psi }}_{{\Omega _T}}^{\rm{H}}{\bf{R}}_{}^{k - 1}} \right\|_F}\\
~~~~~~~~~~~~\mathop  \ge \limits^{}  {\left. {\left\| {{\bf{\Psi }}_{{\Omega _T}}^{\rm{H}}({\bf{\Psi D}} - {\bf{\Psi }}_{{{\tilde \Omega }^{k - 1}}}^{}{\bf{\Psi }}_{{{\tilde \Omega }^{k - 1}}}^\dag {\bf{\Psi D}})} \right.} \right\|_F}\\ ~~~~~~~~~~~~~~~~~~~~~~~
- {\left. {\left\| {{\bf{\Psi }}_{{\Omega _T}}^{\rm{H}}({\bf{W}} - {\bf{\Psi }}_{{{\tilde \Omega }^{k - 1}}}^{}{\bf{\Psi }}_{{{\tilde \Omega }^{k - 1}}}^\dag {\bf{W}})} \right.} \right\|_F}\\
~~~~~~~~~~~~\mathop  \ge \limits^{}  {\left. {\left\| {{\bf{\Psi }}_{{\Omega _T}}^{\rm{H}}{\bf{R'}}_{}^{k - 1}} \right.} \right\|_F}-\sqrt {1 + {\delta _P}} {\left\| {\bf{W}} \right\|_F}.
\end{array}
 \end{small}
\end{equation}
%}
Combining (\ref{equ:app3}) and (\ref{equ:app3.1}), we have
%{\setlength\abovedisplayskip{6pt}
%\setlength\belowdisplayskip{6pt}
 \begin{equation}\label{equ:app3.2}
 \begin{small}
\begin{array}{l}
\!\!\!\!\!\!\!{\left\| {{\bf{\Psi }}_{{\Omega _\Delta }}^{\rm{H}}} {\bf{R'}}_{}^{k - 1} \right\|_F}
\ge   {\left. {\left\| {{\bf{\Psi }}_{{\Omega _T}}^{\rm{H}}{\bf{R'}}_{}^{k - 1}} \right.} \right\|_F}-2\sqrt {1 + {\delta _P}} {\left\| {\bf{W}} \right\|_F}.
\end{array}
 \end{small}
\end{equation}%}
%Then we investigate ${\left. {\left\| {{\bf{\Psi }}_{{\Omega _T}}^{\rm{H}}{\bf{R'}}_{}^{k - 1}} \right.} \right\|_F}$, and we can obtain
%{\setlength\abovedisplayskip{6pt}
%\setlength\belowdisplayskip{6pt}
Due to the following inequality
\begin{equation}\label{equ:app3.23}
\begin{small}
{\left\| {{\bf{\Psi }}_{{\Omega _\Delta }}^{\rm{H}}} {\bf{R'}}_{}^{k - 1} \right\|_F}\ge {\left\| {{\bf{\Psi }}_{{\Omega _T}}^{\rm{H}}{\bf{R}}_{}^{'k - 1}} \right\|_F} \ge {\left\| {{\bf{\Psi }}_{{\Omega _T}/{\tilde \Omega ^{k - 1}}}^{\rm{H}}{\bf{R}}_{}^{'k - 1}} \right\|_F},
\end{small}
\end{equation}%}
(\ref{equ:app3.2}) can be further expressed as the following inequality by removing the common set of ${\Omega _\Delta }$ and ${ \Omega _T}/{\tilde \Omega ^{k - 1}}$, i.e.,
 \begin{equation}\label{equ:app3.21}
 \begin{small}
\begin{array}{l}
\!\!\!\!\!\!{\left\| {{\bf{\Psi }}_{{\Omega _\Delta }/{\Omega _T}}^{\rm{H}}{\bf{R}}_{}^{'k - 1}} \right\|_F}
\ge   {\left\| {{\bf{\Psi }}_{\{{\Omega _T}/{\tilde \Omega ^{k - 1}}\}/{\Omega _\Delta }}^{\rm{H}}{\bf{R}}_{}^{'k - 1}} \right\|_F}-2\sqrt {1 + {\delta _P}} {\left\| {\bf{W}} \right\|_F},
\end{array}
 \end{small}
\end{equation}
here ${\left\| {{\bf{\Psi }}_{\{{\Omega _T}/{\tilde \Omega ^{k - 1}}\}/{\Omega _\Delta }}^{\rm{H}}{\bf{R}}_{}^{'k - 1}} \right\|_F}$ can be expressed as
%{\setlength\abovedisplayskip{6pt}
%\setlength\belowdisplayskip{6pt}
\begin{equation}\label{equ:app3.3}
\begin{small}
\begin{array}{l}
\!\!\!\!\!\!\!\!\!\!\!\!\!\!{\left\| {{\bf{\Psi }}_{\{{\Omega _T}/{\tilde \Omega ^{k - 1}}\}/{\Omega _\Delta }}^{\rm{H}}{\bf{R}}_{}^{'k - 1}} \right\|_F}
 = {\left\| {{\bf{\Psi }}_{{\Omega _T}/{{\tilde \Omega }^{'k}}}^{\rm{H}}{\bf{R}}_{}^{'k - 1}} \right\|_F}\\
 ={\left\| {{\bf{\Psi }}_{{\Omega _T}/{{\tilde \Omega }^{'k}}}^{\rm{H}}{{\bf{\Psi }}_{{\Omega _T} \cup {{\tilde \Omega }^{k - 1}}}}{\bf{\tilde D}}_{}^{k - 1}} \right\|_F}\\
 = \left\| {{\bf{\Psi }}_{{\Omega _T}/{{\tilde \Omega }^{'k}}}^{\rm{H}}({{\bf{\Psi }}_{\{{\Omega _T} \cup {{\tilde \Omega }^{k - 1}}\} /\{ {\Omega _T}/{{\tilde \Omega }^{'k}}\}}}{\bf{\tilde D}}_{ \{ {\Omega _T} \cup {{\tilde \Omega }^{k - 1}} \}/\{ {\Omega _T}/{{\tilde \Omega }^{'k}}\}}^{k - 1}} \right.\\~~~~~~~~~~~~~~
+ {\left. {{{\bf{\Psi }}_{{\Omega _T}/{{\tilde \Omega }^{'k}}}}{\bf{\tilde D}}_{{\Omega _T}/{{\tilde \Omega }^{'k}}}^{k - 1})} \right\|_F}\\
\ge {\left. {\left\| {{{\bf{\Psi }}^{\rm{H}}_{{\Omega _T}/{{\tilde \Omega }^{'k}}}}{{\bf{\Psi }}_{{\Omega _T}/{{\tilde \Omega }^{'k}}}}{\bf{\tilde D}}_{{\Omega _T}/{{\tilde \Omega }^{'k}}}^{k - 1}} \right.} \right\|_F}\\~~~~
- {\left. {\left\| {{\bf{\Psi }}_{{\Omega _T}/{{\tilde \Omega }^{'k}}}^{\rm{H}}{{\bf{\Psi }}_{\{{\Omega _T} \cup {{\tilde \Omega }^{k - 1}}\} / \{{\Omega _T}/{{\tilde \Omega }^{'k}}\}}}{\bf{\tilde D}}_{\{{\Omega _T} \cup {{\tilde \Omega }^{k - 1}} \}/\{ {\Omega _T}/{{\tilde \Omega }^{'k}}\}}^{k - 1}} \right.} \right\|_F}\\
 \ge (1 - {\delta _P}){\left\| {{\bf{\tilde D}}_{{\Omega _T}/{{\tilde \Omega }^{'k}}}^{k - 1}} \right\|_F} - {\delta _{3P}}{\left\| {{\bf{\tilde D}}_{}^{k - 1}} \right\|_F}\\
= (1 - {\delta _P}){\left\| {{\bf{ D}}_{{\Omega _T}/{{\tilde \Omega }^{'k}}}} \right\|_F} - {\delta _{3P}}{\left\| {{\bf{\tilde D}}_{}^{k - 1}} \right\|_F},
\end{array}
\end{small}
\end{equation}
%}
where the first equality is due to ${\Omega _\Delta } \cap {{\tilde \Omega }^{k - 1}}= \phi $ and ${\Omega _\Delta } \cup {\tilde \Omega ^{k - 1}} = {\tilde \Omega '^k}$, the second equality is due to (\ref{equ:app2}), and the last equality is due to the definition of ${\bf{\tilde D}}_{}^{k - 1}$.
Since ${\left\| {{\bf{\Psi }}_{{\Omega _\Delta }/{\Omega _T}}^{\rm{H}}{\bf{R}}_{}^{'k - 1}} \right\|_F}={\left\| {{\bf{\Psi }}_{{\Omega _\Delta }/{\Omega _T}}^{\rm{H}}{{\bf{\Psi }}_{{\Omega _T} \cup {{\tilde \Omega }^{k - 1}}}}{\bf{\tilde D}}_{}^{k - 1}} \right\|_F}\le {\delta _{3P}}{\left\| {{\bf{\tilde D}}_{}^{k - 1}} \right\|_F}$, by substituting (\ref{equ:app3.3}) into (\ref{equ:app3.21}), we have
%{\setlength\abovedisplayskip{6pt}
%\setlength\belowdisplayskip{6pt}
\begin{equation}\label{equ:app4}
\begin{array}{l}
(1 - {\delta _P}){\left\| {{\bf{ D}}_{{\Omega _T}/{{\tilde \Omega }^{'k}}}^{}} \right\|_F} \le 2{\delta _{3P}}{\left\| {{\bf{\tilde D}}_{}^{k - 1}} \right\|_F} + 2\sqrt {1 + {\delta _P}} {\left\| {\bf{W}} \right\|_F}.
\end{array}
\end{equation}%}
It should be pointed out that for ${\left\| {{\bf{\tilde D}}_{}^{k - 1}} \right\|_F} $, we can further get
%{\setlength\abovedisplayskip{6pt}
%\setlength\belowdisplayskip{6pt}
\begin{equation}\label{equ:app5}
\begin{small}
\begin{array}{l}
{\left\| {{\bf{\tilde D}}_{}^{k - 1}} \right\|_F} \le {\left\| {{\bf{D}}_{{\Omega _T}/{{\tilde \Omega }^{k - 1}}}^{}} \right\|_F} + {\left\| {{\bf{\Psi }}_{{{\tilde \Omega }^{k - 1}}}^\dag {{\bf{\Psi }}_{{\Omega _T}/{{\tilde \Omega }^{k - 1}}}}{{\bf{D}}_{{\Omega _T}/{{\tilde \Omega }^{k - 1}}}}} \right\|_F}\\
~~~~~~~~~~~~ = {\left\| {{\bf{D}}_{{\Omega _T}/{{\tilde \Omega }^{k - 1}}}^{}} \right\|_F}\\ ~~~~~~~~~~~~~~
  + {\left\| {{{({\bf{\Psi }}_{{{\tilde \Omega }^{k - 1}}}^{\rm{H}}{\bf{\Psi }}_{{{\tilde \Omega }^{k - 1}}}^{})}^{ - 1}}{\bf{\Psi }}_{{{\tilde \Omega }^{k - 1}}}^{\rm{H}}{{\bf{\Psi }}_{{\Omega _T}/{{\tilde \Omega }^{k - 1}}}}{{\bf{D}}_{{\Omega _T}/{{\tilde \Omega }^{k - 1}}}}} \right\|_F}\\
 ~~~~~~~~~~~~ \le {\left\| {{\bf{D}}_{{\Omega _T}/{{\tilde \Omega }^{k - 1}}}^{}} \right\|_F} + \frac{{{\delta _{2P}}}}{{1 - {\delta _P}}}{\left\| {{{\bf{D}}_{{\Omega _T}/{{\tilde \Omega }^{k - 1}}}}} \right\|_F}\\
~~~~~~~~~~~~   =\frac{1-{\delta _{P}}+{\delta _{2P}}}{{1 - {\delta _{P}}}}{\left\| {{{\bf{D}}_{{\Omega _T}/{{\tilde \Omega }^{k - 1}}}}} \right\|_F},
\end{array}
\end{small}
\end{equation}%}
where the first inequality is due to the definition of ${\bf{\tilde D}}_{}^{k - 1}$. By substituting (\ref{equ:app4}) into (\ref{equ:app5}), we have
\vspace*{-2.0mm}
\begin{equation}\label{equ:app6}
\begin{small}
\begin{array}{l}
{\left\| {{{\bf{D}}_{{\Omega _T}/{{\tilde \Omega }^{k - 1}}}}} \right\|_F} \ge \frac{{{{(1 - {\delta _P})}^2}}}{{2{\delta _{3P}}(1{\rm{ - }}{\delta _P}{\rm{ + }}{\delta _{2P}})}}{\left\| {{{\bf{D}}_{{\Omega _T}/{{\tilde \Omega }^{'k}}}}} \right\|_F}\\~~~~~~~~~~~~~~ ~~~~~~~~~
 - \frac{{\sqrt {1 + {\delta _P}} (1{\rm{ - }}{\delta _P})}}{{{\delta _{3P}}(1{\rm{ - }}{\delta _P}{\rm{ + }}{\delta _{2P}})}}{\left\| {\bf{W}} \right\|_F}.
\end{array}
\end{small}
\end{equation}
Then, we investigate ${{{\bf{D}}_{{\Omega _T}/{{\tilde \Omega }^{k }}}}}$, which can be expressed as
\begin{equation}\label{equ:app7}
\begin{small}\begin{array}{l}
{\left\| {{{\bf{D}}_{{\Omega _T}/{{\tilde \Omega }^k}}}} \right\|_F} = {\left\| {{{\bf{D}}_{{\Omega _T} \cap \{ {{\tilde \Omega }^{'k}}/{{\tilde \Omega }^k}\}  + {\Omega _T}/{{\tilde \Omega }^{'k}}}}} \right\|_F}\\ ~~~~~~~~~~~~~~~
\le {\left\| {{{\bf{D}}_{{\Omega _T} \cap \{ {{\tilde \Omega}^{'k}}/{{\tilde \Omega }^k}\} }}} \right\|_F} + {\left\| {{{\bf{D}}_{{\Omega _T}/{{\tilde \Omega}^{'k}}}}} \right\|_F}\\ ~~~~~~~~~~~~~~~= {\left\| {{{\bf{D}}_{{{\tilde \Omega }^{'k}}/{{\tilde \Omega }^k}}}} \right\|_F} + {\left\| {{{\bf{D}}_{{\Omega _T}/{{\tilde \Omega }^{'k}}}}} \right\|_F},
\end{array}
\end{small}
\end{equation}
where we use the fact that ${{\tilde \Omega }^{k}}  \subset  {{\tilde \Omega }^{'k}}$.
For ${\left\| {{{\bf{D}}_{{{\tilde \Omega }^{'k}}/{{\tilde \Omega }^k}}}} \right\|_F} $, we can further obtain
\begin{equation}\label{equ:app8}
\begin{small}
\begin{array}{l}
{\left\| {{{\bf{D}}_{{{\tilde \Omega }^{'k}}/{{\tilde \Omega }^k}}}} \right\|_F} = {\left\| {{{{{{\bf{\mathord{\buildrel{\lower3pt\hbox{$\scriptscriptstyle\smile$}}
\over D}
}}}_{{{\tilde \Omega }^{'k}}\cap \{{{{\tilde \Omega }^{'k}}/{{\tilde \Omega }^k}}\}}} + {\bf{E}}}_{{{\tilde \Omega }^{'k}}/{{\tilde \Omega }^k}}}} \right\|_F}\\~~~~~~~~~~~~~~~~
\le {\left\| {{{\bf{\mathord{\buildrel{\lower3pt\hbox{$\scriptscriptstyle\smile$}}
\over D}
}}}_{{{\tilde \Omega }^{'k}}\cap \{{{{\tilde \Omega }^{'k}}/{{\tilde \Omega }^k}}\}}}  \right\|_F} + {\left\| {{{\bf{E}}_{{{\tilde \Omega }^{'k}}/{{\tilde \Omega }^k}}}} \right\|_F}\\~~~~~~~~~~~~~~~~
 \le {\left\| {{{\bf{\mathord{\buildrel{\lower3pt\hbox{$\scriptscriptstyle\smile$}}
\over D}
}}}_{{{\tilde \Omega }^{'k}}\cap {\Omega '}}} \right\|_F} + {\left\| {{{\bf{E}}_{{{\tilde \Omega }^{'k}}/{{\tilde \Omega }^k}}}} \right\|_F}\\~~~~~~~~~~~~~~~~
  = {\left\| {{{{\bf{D}}_{{{\tilde \Omega }^{'k}}\cap {\Omega '}} - {\bf{E}}}_{\Omega '}}} \right\|_F} + {\left\| {{{\bf{E}}_{{{\tilde \Omega }^{'k}}/{{\tilde \Omega }^k}}}} \right\|_F}\\~~~~~~~~~~~~~~~~
 \le {\left\| {{{\bf{D}}_{\Omega '}}} \right\|_F} + {\left\| {{{\bf{E}}_{\Omega '}}} \right\|_F} + {\left\| {{{\bf{E}}_{{{\tilde \Omega }^{'k}}/{{\tilde \Omega }^k}}}} \right\|_F}\\
~~~~~~~~~~~~~~~~ = {\bf{0}}+{\left\| {{{\bf{E}}_{\Omega '}}} \right\|_F} + {\left\| {{{\bf{E}}_{{{\tilde \Omega }^{'k}}/{{\tilde \Omega }^k}}}} \right\|_F}\\
~~~~~~~~~~~~~~~~
\le 2{\left\| {\bf{E}} \right\|_F},
\end{array}
\end{small}
\end{equation}
where we introduce the error variable ${\bf{E}}=  {\bf{D}}_{{{\tilde \Omega }^{'k}}}-{{{\bf{\mathord{\buildrel{\lower3pt\hbox{$\scriptscriptstyle\smile$}}
\over D}
}}}_{{{\tilde \Omega }^{'k}}}}$ (${{{\bf{\mathord{\buildrel{\lower3pt\hbox{$\scriptscriptstyle\smile$}}
\over D}
}}}_{{{\tilde \Omega }^{'k}}}}$ is obtained in \textbf{Step 2.3} of Algorithm 1), and $ \Omega'$ is an arbitrary set satisfying ${\left| {\Omega '} \right|_c} = P$, $\Omega ' \subset {{\tilde \Omega }^{'k}}$, and $\Omega ' \cap {\Omega _T} = \phi $. The second inequality in (\ref{equ:app8}) is due to the fact that ${{{\tilde \Omega }^{'k}}/{{\tilde \Omega }^k}}$ is the discarded support in the step of support pruning in Algorithm 1. According to the definition of ${\bf{E}}$, we further obtain
\begin{equation}\label{equ:app9}
\begin{small}
\begin{array}{l}
{\left\| {\bf{E}} \right\|_F} = {\left\| {{{\bf{D}}_{{{\tilde \Omega }^{'k}}}} - {{{\bf{\mathord{\buildrel{\lower3pt\hbox{$\scriptscriptstyle\smile$}}
\over D}
}}}_{{{\tilde \Omega }^{'k}}}}} \right\|_F}{\rm{ = }}{\left\| {{{\bf{D}}_{{{\tilde \Omega }^{'k}}}} - {\bf{\Psi }}_{{{\tilde \Omega }^{'k}}}^\dag {\bf{Y}}} \right\|_F}\\~~~~~~~~
{\rm{ = }}{\left\| {{{\bf{D}}_{{{\tilde \Omega }^{'k}}}} - {\bf{\Psi }}_{{{\tilde \Omega }^{'k}}}^\dag ({\bf{\Psi D}} + {\bf{W}})} \right\|_F}\\
~~~~~~~\le {\left\| {{{\bf{D}}_{{{\tilde \Omega }^{'k}}}} - {\bf{\Psi }}_{{{\tilde \Omega }^{'k}}}^\dag {\bf{\Psi D}}} \right\|_F}{\rm{ + }}{\left\| {{\bf{\Psi }}_{{{\tilde \Omega }^{'k}}}^\dag {\bf{W}}} \right\|_F}
\\~~~~~~~
= {\left\| {{{\bf{D}}_{{{\tilde \Omega }^{'k}}}} - {\bf{\Psi }}_{{{\tilde \Omega }^{'k}}}^\dag {\bf{\Psi}}_{\Omega_T} {{\bf D}}_{\Omega_T}} \right\|_F}{\rm{ + }}{\left\| {{\bf{\Psi }}_{{{\tilde \Omega }^{'k}}}^\dag {\bf{W}}} \right\|_F}.
\end{array}
\end{small}
\end{equation}
For ${\left\| {{{\bf{D}}_{{{\tilde \Omega }^{'k}}}} - {\bf{\Psi }}_{{{\tilde \Omega }^{'k}}}^\dag {{\bf{\Psi }}_{{\Omega _T}}}{{\bf{D}}_{{\Omega _T}}}} \right\|_F}$, we can have
\begin{equation}\label{equ:app10}
\begin{small}
\begin{array}{l}
{\left\| {{{\bf{D}}_{{{\tilde \Omega }^{'k}}}} - {\bf{\Psi }}_{{{\tilde \Omega }^{'k}}}^\dag {{\bf{\Psi }}_{{\Omega _T}}}{{\bf{D}}_{{\Omega _T}}}} \right\|_F}\\
= {\left\| {{{\bf{D}}_{{{\tilde \Omega }^{'k}}}} - {\bf{\Psi }}_{{{\tilde \Omega }^{'k}}}^\dag ({{\bf{\Psi }}_{{\Omega _T} \cap {{\tilde \Omega }^{'k}}}}{{\bf{D}}_{{\Omega _T} \cap {{\tilde \Omega }^{'k}}}} + {{\bf{\Psi }}_{{\Omega _T}/{{\tilde \Omega }^{'k}}}}{{\bf{D}}_{{\Omega _T}/{{\tilde \Omega }^{'k}}}})} \right\|_F}\\
 %~~~~~~~~~ ~~~~~~ ~~~~~~~~~~ ~~~~~~~~
  = {\left\| {({{\bf{D}}_{{{\tilde \Omega }^{'k}}}}\!\! - \!\! {\bf{\Psi }}_{{{\tilde \Omega }^{'k}}}^\dag {{\bf{\Psi }}_{{\Omega _T} \cap {{\tilde \Omega }^{'k}}}}{{\bf{D}}_{{\Omega _T} \cap {{\tilde \Omega }^{'k}}}}) \!\!- \!\!{\bf{\Psi }}_{{{\tilde \Omega }^{'k}}}^\dag {{\bf{\Psi }}_{{\Omega _T}/{{\tilde \Omega }^{'k}}}}{{\bf{D}}_{{\Omega _T}/{{\tilde \Omega }^{'k}}}}}\! \right\|_F}\\
 %~~~~~~~~~ ~~~~~~ ~~~~~~~~~~ ~~~~~~~~
 = {\left\| {({{\bf{D}}_{{{\tilde \Omega }^{'k}}}} - {\bf{\Psi }}_{{{\tilde \Omega }^{'k}}}^\dag {{\bf{\Psi }}_{{{\tilde \Omega }^{'k}}}}{{\bf{D}}_{{{\tilde \Omega }^{'k}}}}) - {\bf{\Psi }}_{{{\tilde \Omega }^{'k}}}^\dag {{\bf{\Psi }}_{{\Omega _T}/{{\tilde \Omega }^{'k}}}}{{\bf{D}}_{{\Omega _T}/{{\tilde \Omega }^{'k}}}}} \right\|_F}\\
 %~~~~~~~~~ ~~~~~~ ~~~~~~~~~~ ~~~~~~~~
 = {\left\| {{{\bf{D}}_{{{\tilde \Omega }^{'k}}}} - {{\bf{D}}_{{{\tilde \Omega }^{'k}}}} - {\bf{\Psi }}_{{{\tilde \Omega }^{'k}}}^\dag {{\bf{\Psi }}_{{\Omega _T}/{{\tilde \Omega }^{'k}}}}{{\bf{D}}_{{\Omega _T}/{{\tilde \Omega }^{'k}}}}} \right\|_F}\\
 %~~~~~~~~ ~~~~~~~ ~~~~~~~~~~ ~~~~~~~~
  = {\left\| {{\bf{\Psi }}_{{{\tilde \Omega }^{'k}}}^\dag {{\bf{\Psi }}_{{\Omega _T}/{{\tilde \Omega }^{'k}}}}{{\bf{D}}_{{\Omega _T}/{{\tilde \Omega }^{'k}}}}} \right\|_F}\\
 %~~~~~~~~ ~~~~~~~ ~~~~~~~~~~ ~~~~~~~~
 \le \frac{{{\delta _{3P}}}}{{1{\rm{ - }}{\delta _{2P}}}}{\left\| {{{\bf{D}}_{{\Omega _T}/{{\tilde \Omega }^{'k}}}}} \right\|_F},
\end{array}
\end{small}
\end{equation}
where the last inequality is due to $|{{\tilde \Omega }^{'k}}|_c = 2P$. While for ${\left\| {{\bf{\Psi }}_{{{\tilde \Omega }^{'k}}}^\dag {\bf{W}}} \right\|_F}$ in (\ref{equ:app9}), we have
\begin{equation}\label{equ:app11}
\begin{small}
{\left\| {{\bf{\Psi }}_{{{\tilde \Omega }^{'k}}}^\dag {\bf{W}}} \right\|_F} \le {\delta _{2P}}/\sqrt {1{\rm{ - }}{\delta _{2P}}} {\left\| {\bf{W}} \right\|_F}.
\end{small}
\end{equation}
By substituting (\ref{equ:app8})-(\ref{equ:app11}) into (\ref{equ:app7}), we can obtain
\begin{equation}\label{equ:app12}
\begin{small}
{\left\| {{{\bf{D}}_{{\Omega _T}/{{\tilde \Omega }^{'k}}}}} \right\|_F} \ge \frac{{(1{\rm{ - }}{\delta _{2P}}){{\left\| {{{\bf{D}}_{{\Omega _T}/{{\tilde \Omega }^k}}}} \right\|}_F} - 2{\delta _P}\sqrt {1{\rm{ - }}{\delta _{2P}}} {{\left\| {\bf{W}} \right\|}_F}}}{{1{\rm{ - }}{\delta _{2P}} + 2{\delta _{3P}}}}.
\end{small}
\end{equation}
Furthermore, by substituting (\ref{equ:app12}) into (\ref{equ:app6}), we can obtain
\begin{equation}\label{equ:app12.1}
\begin{small}
\begin{array}{*{20}{l}}
\!\!\!\!\!{{{\left\| {{{\bf{D}}_{{\Omega _T}/{{\tilde \Omega }^{k - 1}}}}} \right\|}_F} \ge \underbrace {\frac{{{{(1 - {\delta _P})}^2}(1 - {\delta _{2P}})}}{{2{\delta _{3P}}(1 - {\delta _P} + {\delta _{2P}})(1 - {\delta _{2P}}+{\delta _{3P}})}}}_{{C_1}}{{\left\| {{{\bf{D}}_{{\Omega _T}/{{\tilde \Omega }^k}}}} \right\|}_F}}\\
\!\!\!\!\! - \underbrace {\frac{{(1 - {\delta _P})}}{{{\delta _{3P}}(1 - {\delta _P} + {\delta _{2P}})}}(\frac{{{\delta _P}(1 - {\delta _P})\sqrt {1 - {\delta _{2P}}} }}{{(1 - {\delta _{2P}} + 2{\delta _{3P}})}} + \sqrt {1 + {\delta _P}} )}_{{C_2}}{\left\| {\bf{W}} \right\|_F}.
\end{array}
\end{small}
\end{equation}
As we have discussed, if ${\left\| {{{\bf{R}}^{k - 1}}} \right\|_F} \le {\left\| {\bf{R}}^{k} \right\|_F}$, the iteration quits, which indicates that the estimation of the $P$-sparse signal $\bf{D}$ is obtained, and $\hat \Omega = {{\tilde \Omega }^{k-1 }}$. Then we can combine (\ref{equ:app0.1}), (\ref{equ:app0.2}), and (\ref{equ:app12.1}) to obtain
\begin{equation}\label{equ:14}
\begin{small}
{\left\| {{{\bf{D}}_{{\Omega _T}/\hat \Omega }}} \right\|_F} \le C_3{\left\| {\bf{W}} \right\|_F},
\end{small}
\end{equation}
where ${C_3} = \frac{{2{C_1}\sqrt {1 - \delta _P^{}}  + {C_2}\sqrt {1 - \delta _P^2} }}{{{C_1}(1 - {\delta _P} - {\delta _{2P}}) - \sqrt {1 - \delta _P^2} }}$.
By substituting (\ref{equ:16.1}) into (\ref{equ:14}), we have

\begin{equation}\label{equ:15}
\begin{small}
{\left\| {{\bf{D}} - {\bf{\hat D}}} \right\|_F}\le C_4{\left\| {\bf{W}} \right\|_F},
\end{small}
\end{equation}
where ${C_4} = \frac{{{C_3}(1 - {\delta _P} + {\delta _{2P}}) + \sqrt {1 + \delta _P^{}} }}{{1 - \delta _P^{}}}$. Thus we prove (\ref{equ:app001}).
Finally, in the iterative process, we have ${\left\| {{{\bf{R}}^{k - 1}}} \right\|_F} > {\left\| {\bf{R}}^{k} \right\|_F}$, and by substituting (\ref{equ:app0.1}) and (\ref{equ:app0.2}) into (\ref{equ:app12.1}), we can obtain
\begin{equation}\label{equ:18}
\begin{small}
\begin{array}{l}
{\left\| {{\bf{R}}_{}^{k - 1}} \right\|_F} > \frac{{{C_1}(1 - {\delta _P} - {\delta _{2P}})}}{{\sqrt {1 - \delta _P^2} }}{\left\| {{\bf{R}}_{}^k} \right\|_F}\\~~~~~~~~~~~~~~~~~
 - (1 + \frac{{(1 - {\delta _P} - {\delta _{2P}})({C_1} + {C_2}\sqrt {1 + \delta _P^{}} )}}{{\sqrt {1 - \delta _P^2} }}){\left\| {\bf{W}} \right\|_F}.
\end{array}
\end{small}
\end{equation}
In this way, we prove (\ref{equ:app002}).
\vspace*{-4.0mm}
\subsection{Proof of (\ref{equ:app0031})}\label{TPb}
We consider two matrices ${\bf{D}}'$ and ${\bf{D}}$ have the structured sparsity as illustrated in (\ref{equ:common3}), and both of them have the respective structured support set ${\Omega_1}$ and ${\Omega_1}$, where ${\Omega _1} \cap {\Omega _2} = \phi $. Moreover, we consider ${\bf{\bar D'}} = {\bf{D'}}/{\left\| {{\bf{D'}}} \right\|_F}$ and ${\bf{\bar D}} = {\bf{D}}/{\left\| {\bf{D}} \right\|_F}$. According to (\ref{equ:app003}), we can obtain
\begin{equation}\label{equ:19}
\begin{small}
\begin{array}{l}
\!\!\!\!2(1 - {\delta _{{{\left| {{\Omega _1}} \right|}_c} + {{\left| {{\Omega _2}} \right|}_c}}}) \le {\left\| {[{{\bf{\Psi }}_{{\Omega _1}}},{{\bf{\Psi }}_{{\Omega _2}}}]\left[\!\! {\begin{array}{*{20}{c}}
{{{{\bf{\bar D'}}}_{{\Omega _1}}}}\\
{{{{\bf{\bar D}}}_{{\Omega _2}}}}
\end{array}}\!\!\! \right]}\! \right\|_F^2}\!\!\!\! \le 2(1 + {\delta _{{{\left| {{\Omega _1}} \right|}_c} + {{\left| {{\Omega _2}} \right|}_c}}}\!),
\end{array}
\end{small}
\end{equation}
%\begin{equation}\label{equ:19}
%\begin{small}
%\begin{array}{l}
%\!\!\!\!\!\!\!2(1 - {\delta _{{{\left| {{\Omega _1}} \right|}_c} + {{\left| {{\Omega _2}} \right|}_c}}}) \le {\left\| {[{{\bf{\Psi }}_{{\Omega %_1}}},{{\bf{\Psi }}_{{\Omega _2}}}]\left[ {\begin{array}{*{20}{c}}
%{{{{\bf{\bar D'}}}_{{\Omega _1}}}}\\
%{{{{\bf{\bar D}}}_{{\Omega _2}}}}
%\end{array}} \right]} \right\|_F^2}\\
%  ~~~~~~~~~~~~~~~~~~~~~~~~~~~~~~~~~~~\le 2(1 + {\delta _{{{\left| {{\Omega _1}} \right|}_c} + {{\left| {{\Omega _2}} \right|}_c}}}),
%\end{array}
%\end{small}
%\end{equation}
\begin{equation}\label{equ:20}
\begin{small}
\begin{array}{l}
\!\!\!\!\!\!\!2(1 - {\delta _{{{\left| {{\Omega _1}} \right|}_c} + {{\left| {{\Omega _2}} \right|}_c}}}) \le {\left\| {[\!{{\bf{\Psi }}_{{\Omega _1}}},{{\bf{\Psi }}_{{\Omega _2}}}\!]\left[\!\! {\begin{array}{*{20}{c}}
{{{{\bf{\bar D'}}}_{{\Omega _1}}}}\\
{{{-{\bf{\bar D}}}_{{\Omega _2}}}}
\end{array}}\!\! \right]} \! \right\|_F^2}%\\  ~~~~~~~~~~~~~~~~~~~~~~~~~~~~~~~~~~~
\!\!\!  \le \!\!2(1 + {\delta _{{{\left| {{\Omega _1}} \right|}_c} + {{\left| {{\Omega _2}} \right|}_c}}}).
\end{array}
\end{small}
\end{equation}
From (\ref{equ:19}) and (\ref{equ:20}), we obtain
\begin{equation}\label{equ:21}
\begin{small}
{\rm{ - }}{\delta _{{{\left| {{\Omega _1}} \right|}_c} + {{\left| {{\Omega _2}} \right|}_c}}} \le {\mathop{\rm Re}\nolimits} {\rm{\{ }}\left\langle {{{\bf{\Psi }}_{{\Omega _1}}}{{{\bf{\bar D'}}}_{{\Omega _1}}},{{\bf{\Psi }}_{{\Omega _2}}}{{{\bf{\bar D}}}_{{\Omega _2}}}} \right\rangle {\rm{\} }} \le {\delta _{{{\left| {{\Omega _1}} \right|}_c} + {{\left| {{\Omega _2}} \right|}_c}}},
\end{small}
\end{equation}
where for two matrices $\bf{A}$ and $\bf{B}$, we have ${\mathop{\rm Re}\nolimits} {\rm{\{ }}\left\langle {{\bf{A}},{\bf{B}}} \right\rangle {\rm{\} }} = \frac{{\left\| {{\bf{A}}{\rm{ + }}{\bf{B}}} \right\|_F^2{\rm{ - }}\left\| {{\bf{A}}{\rm{ - }}{\bf{B}}} \right\|_F^2}}{4}$. Moreover, we exploit the Cauchy-Schwartz inequality ${\left\| {\bf{A}} \right\|_F}{\left\| {\bf{B}} \right\|_F} \ge {\left| {\left\langle {{\bf{A}}{\rm{,}}{\bf{B}}} \right\rangle } \right|}$, where the equality holds only for ${\bf{A}}=c{\bf{B}}$ and $c$ is a complex constant. Particularly,
\begin{equation}\label{equ:22}
\begin{small}
\begin{array}{l}
\left\| {{\bf{\bar D}}_{{\Omega _1}}^{'}} \right\|_F^{}\left\| {{\bf{\Psi }}_{{\Omega _1}}^{\rm{H}}{{\bf{\Psi }}_{{\Omega _2}}}{{{\bf{\bar D}}}_{{\Omega _2}}}} \right\|_F^{}\\
 = \mathop {\max }\limits_{{{{\bf{\bar D}}}^{'}_{{\Omega _1}}}{\rm{ = c'}}{\bf{\Psi }}_{{\Omega _1}}^{\rm{H}}{{\bf{\Psi }}_{{\Omega _2}}}{{{\bf{\bar D}}}_{{\Omega _2}}}} \left| {\left\langle {{{\bf{\Psi }}_{{\Omega _1}}}{{{\bf{\bar D}}}^{'}_{{\Omega _1}}}{\rm{,}}{{\bf{\Psi }}_{{\Omega _2}}}{{{\bf{\bar D}}}_{{\Omega _2}}}} \right\rangle } \right|\\
 %~ ~ ~ ~ ~ ~ ~ ~ ~ ~ ~ ~ ~ ~ ~ ~ ~ ~ ~ ~ ~ ~ ~ ~ ~ ~ ~ ~ ~ ~ ~ ~
 = \mathop {\max }\limits_{{{{{\bf{\bar D}}}^{'}_{{\Omega _1}}}{\rm{ = c'}}{\bf{\Psi }}_{{\Omega _1}}^{\rm{H}}{{\bf{\Psi }}_{{\Omega _2}}}{{{\bf{\bar D}}}_{{\Omega _2}}}}} (\left| {\rm{Re}\{ \left\langle {{{\bf{\Psi }}_{{\Omega _1}}}{{{\bf{\bar D}}}^{'}_{{\Omega _1}}}{\rm{,}}{{\bf{\Psi }}_{{\Omega _2}}}{{{\bf{\bar D}}}_{{\Omega _2}}}} \right\rangle \} } \right|)\\ %~ ~ ~ ~ ~ ~ ~ ~ ~ ~ ~ ~ ~ ~ ~ ~ ~ ~ ~ ~ ~ ~ ~ ~ ~ ~ ~ ~ ~ ~ ~ ~
 \le {\delta _{{{\left| {{\Omega _1}} \right|}_c} + {{\left| {{\Omega _2}} \right|}_c}}},
\end{array}
\end{small}
\end{equation}
where $c'$ is a complex constant, and the second equality of (\ref{equ:22}) is due to ${\rm{Im}}\{ \left\langle {{{\bf{\Psi }}_{{\Omega _1}}}{{{\bf{\bar D'}}}_{{\Omega _1}}},{{\bf{\Psi }}_{{\Omega _2}}}{{{\bf{\bar D}}}_{{\Omega _2}}}} \right\rangle \}  = c'{\rm{Im}}\{ \left\langle {{\bf{\Psi }}_{{\Omega _1}}^{\rm{H}}{{\bf{\Psi }}_{{\Omega _2}}}{{{\bf{\bar D}}}_{{\Omega _2}}},{\bf{\Psi }}_{{\Omega _1}}^{\rm{H}}{{\bf{\Psi }}_{{\Omega _2}}}{{{\bf{\bar D}}}_{{\Omega _2}}}} \right\rangle \} {\rm{ = }}0$. In this way, we have
\begin{equation}\label{equ:23}
\begin{small}
\left\| {{\bf{\Psi }}_{{\Omega _1}}^{\rm{H}}{{\bf{\Psi }}_{{\Omega _2}}}{{\bf{D}}_{{\Omega _2}}}} \right\|_F^{}  \le {\delta _{{{\left| {{\Omega _1}} \right|}_c} + {{\left| {{\Omega _2}} \right|}_c}}}{\left\| {{{\bf{D}}_{{\Omega _2}}}} \right\|_F},
\end{small}
\vspace*{-2.0mm}
\end{equation}
and (\ref{equ:app0031}) is proven.
\vspace*{-4.0mm}
\subsection{Proof of (\ref{equ:app004})}\label{TPc}
Clearly, we have
\begin{equation}\label{equ:24}
\begin{small}
\begin{array}{l}
{\left\| \!{({\bf{I}}{\rm{ - }}{{\bf{\Psi }}_{{\Omega _1}}}{\bf{\Psi }}_{{\Omega _1}}^\dag ){{\bf{\Psi }}_{{\Omega _2}}}{{\bf{D}}_{{\Omega _2}}}} \right\|_F}%\\ ~~~~~~~~~~~~~
\!\! \!\ge \!{\left\| \!{{{\bf{\Psi }}_{{\Omega _2}}}{{\bf{D}}_{{\Omega _2}}}}\! \right\|_F}{\rm{\! -\! }}{\left\| {{{\bf{\Psi }}_{{\Omega _1}}}{\bf{\Psi }}_{{\Omega _1}}^\dag {{\bf{\Psi }}_{{\Omega _2}}}{{\bf{D}}_{{\Omega _2}}}}\! \right\|_F},
\end{array}\end{small}
\end{equation}
For ${\left\| {{{\bf{\Psi }}_{{\Omega _1}}}{\bf{\Psi }}_{{\Omega _1}}^\dag {{\bf{\Psi }}_{{\Omega _2}}}{{\bf{D}}_{{\Omega _2}}}} \right\|_F^2}$, we have
\begin{equation}\label{equ:25}
\begin{small}
\begin{array}{l}
{\left\| {{{\bf{\Psi }}_{{\Omega _1}}}{\bf{\Psi }}_{{\Omega _1}}^\dag {{\bf{\Psi }}_{{\Omega _2}}}{{\bf{D}}_{{\Omega _2}}}} \right\|_F^2}{\rm{ = }}\left\langle {{{\bf{\Psi }}_{{\Omega _1}}}{\bf{\Psi }}_{{\Omega _1}}^\dag {{\bf{\Psi }}_{{\Omega _2}}}{{\bf{D}}_{{\Omega _2}}},{{\bf{\Psi }}_{{\Omega _1}}}{\bf{\Psi }}_{{\Omega _1}}^\dag {{\bf{\Psi }}_{{\Omega _2}}}{{\bf{D}}_{{\Omega _2}}}} \right\rangle \\
{\rm{ = Re\{ }}\left\langle {{{\bf{\Psi }}_{{\Omega _1}}}{\bf{\Psi }}_{{\Omega _1}}^\dag {{\bf{\Psi }}_{{\Omega _2}}}{{\bf{D}}_{{\Omega _2}}},{{\bf{\Psi }}_{{\Omega _1}}}{\bf{\Psi }}_{{\Omega _1}}^\dag {{\bf{\Psi }}_{{\Omega _2}}}{{\bf{D}}_{{\Omega _2}}}} \right\rangle {\rm{\} }}\\
{\rm{ = Re\{ }}\left\langle {{{\bf{\Psi }}_{{\Omega _1}}}{\bf{\Psi }}_{{\Omega _1}}^\dag {{\bf{\Psi }}_{{\Omega _2}}}{{\bf{D}}_{{\Omega _2}}},{{\bf{\Psi }}_{{\Omega _1}}}{\bf{\Psi }}_{{\Omega _1}}^\dag {{\bf{\Psi }}_{{\Omega _2}}}{{\bf{D}}_{{\Omega _2}}}} \right.\\~~~~~~~~~~~~~~~~~~~~~~~~
\left. {{\rm{ + }}{{\bf{\Psi }}_{{\Omega _2}}}{{\bf{D}}_{{\Omega _2}}}{\rm{ - }}{{\bf{\Psi }}_{{\Omega _1}}}{\bf{\Psi }}_{{\Omega _1}}^\dag {{\bf{\Psi }}_{{\Omega _2}}}{{\bf{D}}_{{\Omega _2}}}} \right\rangle {\rm{\} }}\\
{\rm{ = Re\{ }}\left\langle {{{\bf{\Psi }}_{{\Omega _1}}}{\bf{\Psi }}_{{\Omega _1}}^\dag {{\bf{\Psi }}_{{\Omega _2}}}{{\bf{D}}_{{\Omega _2}}},{{\bf{\Psi }}_{{\Omega _2}}}{{\bf{D}}_{{\Omega _2}}}} \right\rangle {\rm{\} }}\\
 \le {\delta _{{{\left| {{\Omega _1}} \right|}_c} + {{\left| {{\Omega _2}} \right|}_c}}}{\left\| {{\bf{\Psi }}_{{\Omega _1}}^\dag {{\bf{\Psi }}_{{\Omega _2}}}{{\bf{D}}_{{\Omega _2}}}} \right\|_F}{\left\| {{{\bf{D}}_{{\Omega _2}}}} \right\|_F}\\
 \le {\delta _{{{\left| {{\Omega _1}} \right|}_c} + {{\left| {{\Omega _2}} \right|}_c}}}\frac{{{{\left\| {{{\bf{\Psi }}_{{\Omega _1}}}{\bf{\Psi }}_{{\Omega _1}}^\dag {{\bf{\Psi }}_{{\Omega _2}}}{{\bf{D}}_{{\Omega _2}}}} \right\|}_F}}}{{\sqrt {1{\rm{ - }}{\delta _{{{\left| {{\Omega _1}} \right|}_c}}}} }}\frac{{{{\left\| {{{\bf{\Psi }}_{{\Omega _2}}}{{\bf{D}}_{{\Omega _2}}}} \right\|}_F}}}{{\sqrt {1{\rm{ - }}{\delta _{{{\left| {{\Omega _2}} \right|}_c}}}} }},
\end{array}
\end{small}
\end{equation}
where the first inequality in (\ref{equ:25}) is due to (\ref{equ:21}), and the third equality in (\ref{equ:25}) is due to the following equality,
\begin{equation}\label{equ:25.1}
\begin{small}
\begin{array}{l}
\left\langle {{{\bf{\Psi }}_{{\Omega _1}}}{\bf{\Psi }}_{{\Omega _1}}^\dag {{\bf{\Psi }}_{{\Omega _2}}}{{\bf{D}}_{{\Omega _2}}},{{\bf{\Psi }}_{{\Omega _2}}}{{\bf{D}}_{{\Omega _2}}}{\rm{ - }}{{\bf{\Psi }}_{{\Omega _1}}}{\bf{\Psi }}_{{\Omega _1}}^\dag {{\bf{\Psi }}_{{\Omega _2}}}{{\bf{D}}_{{\Omega _2}}}} \right\rangle \\
{\rm{ = }}{\bf{D}}_{{\Omega _2}}^{\rm{H}}{\bf{\Psi }}_{{\Omega _2}}^{\rm{H}}{({\bf{\Psi }}_{{\Omega _1}}^\dag )^{\rm{H}}}({\bf{\Psi }}_{{\Omega _1}}^{\rm{H}}{{\bf{\Psi }}_{{\Omega _2}}}{{\bf{D}}_{{\Omega _2}}}{\rm{ - }}{\bf{\Psi }}_{{\Omega _1}}^{\rm{H}}{{\bf{\Psi }}_{{\Omega _1}}}{\bf{\Psi }}_{{\Omega _1}}^\dag {{\bf{\Psi }}_{{\Omega _2}}}{{\bf{D}}_{{\Omega _2}}})\\
{\rm{ = }}{\bf{D}}_{{\Omega _2}}^{\rm{H}}{\bf{\Psi }}_{{\Omega _2}}^{\rm{H}}{({\bf{\Psi }}_{{\Omega _1}}^\dag )^{\rm{H}}}({\bf{\Psi }}_{{\Omega _1}}^{\rm{H}}{{\bf{\Psi }}_{{\Omega _2}}}{{\bf{D}}_{{\Omega _2}}}{\rm{ - }}{\bf{\Psi }}_{{\Omega _1}}^{\rm{H}}{{\bf{\Psi }}_{{\Omega _2}}}{{\bf{D}}_{{\Omega _2}}}){\rm{ = }}0.
\end{array}
\end{small}
\end{equation}
Here ${\bf{\Psi }}_{{\Omega _1}}^\dag {\rm{ = (}}{\bf{\Psi }}_{{\Omega _1}}^{\rm{H}}{\bf{\Psi }}_{{\Omega _1}}^{}{{\rm{)}}^{{\rm{ - }}1}}{\bf{\Psi }}_{{\Omega _1}}^{\rm{H}}$. Moreover, (\ref{equ:25}) can be expressed as
\begin{equation}\label{equ:26}
\begin{small}
{\left\| {{{\bf{\Psi }}_{{\Omega _1}}}{\bf{\Psi }}_{{\Omega _1}}^\dag {{\bf{\Psi }}_{{\Omega _2}}}{{\bf{D}}_{{\Omega _2}}}} \right\|_F} \le \frac{{{\delta _{{{\left| {{\Omega _1}} \right|}_c} + {{\left| {{\Omega _2}} \right|}_c}}}}{\left\| {{{\bf{\Psi }}_{{\Omega _2}}}{{\bf{D}}_{{\Omega _2}}}} \right\|_F}}{{\sqrt {(1 - {\delta _{{{\left| {{\Omega _1}} \right|}_c}}})(1 - {\delta _{{{\left| {{\Omega _2}} \right|}_c}}})} }}.
\end{small}
\end{equation}
By substituting (\ref{equ:26}) into (\ref{equ:24}), we have
\begin{equation}\label{equ:27}
\begin{small}
\!\!\!\!\begin{array}{l}
{\left\| \!{({\bf{I}}{\rm{ - }}{{\bf{\Psi }}_{{\Omega _1}}}{\bf{\Psi }}_{{\Omega _1}}^\dag ){{\bf{\Psi }}_{{\Omega _2}}}{{\bf{D}}_{{\Omega _2}}}}\! \right\|_F}% \\~~~~~~~~~
\!\!\!\ge \!\!(1\!\! - \!\!\frac{{{\delta _{{{\left| {{\Omega _1}} \right|}_c} + {{\left| {{\Omega _2}} \right|}_c}}}}}{\sqrt {(1 \!- \!{\delta _{{{\left| {{\Omega _1}} \right|}_c}}})(1\! -\! {\delta _{{{\left| {{\Omega _2}} \right|}_c}}})} }){\left\| {{{\bf{\Psi }}_{{\Omega _2}}}{{\bf{D}}_{{\Omega _2}}}} \!\right\|_F},
\end{array}
\end{small}
\end{equation}
Thus, the right inequality of (\ref{equ:app004}) is proven. Finally, due to (\ref{equ:25.1}), we have
\begin{equation}\label{equ:28}
\begin{small}
\!\!\!\!\begin{array}{l}
\left\| {{{\bf{\Psi }}_{{\Omega _2}}}{{\bf{D}}_{{\Omega _2}}}} \right\|_F^2\!\! =\!\! \left\| {{{\bf{\Psi }}_{{\Omega _1}}}{\bf{\Psi }}_{{\Omega _1}}^\dag {{\bf{\Psi }}_{{\Omega _2}}}{{\bf{D}}_{{\Omega _2}}}}\! \right\|_F^2 \!\!\!%\\~~~~~~~~~~~~~~
+ \!\! \left\| {({\bf{I}}{\rm{\! - \!}}{{\bf{\Psi }}_{{\Omega _1}}}{\bf{\Psi }}_{{\Omega _1}}^\dag ){{\bf{\Psi }}_{{\Omega _2}}}{{\bf{D}}_{{\Omega _2}}}} \!\right\|_F^2,
\end{array}
\end{small}
\end{equation}
which indicates
\begin{equation}\label{equ:29}
\begin{small}
\left\| {{{\bf{\Psi }}_{{\Omega _2}}}{{\bf{D}}_{{\Omega _2}}}} \right\|_F^{} \ge \left\| {({\bf{I}}{\rm{ - }}{{\bf{\Psi }}_{{\Omega _1}}}{\bf{\Psi }}_{{\Omega _1}}^\dag ){{\bf{\Psi }}_{{\Omega _2}}}{{\bf{D}}_{{\Omega _2}}}} \right\|_F^{}.
\end{small}
\end{equation}
Hence the left inequality of (\ref{equ:app004}) is proven.

\end{document}